\documentclass[11pt]{article}
    
\usepackage[bookmarks,colorlinks,breaklinks]{hyperref}  
\hypersetup{linkcolor=blue,citecolor=blue,filecolor=blue,urlcolor=blue} 
\usepackage{fullpage,appendix}
\usepackage{amsmath,amsfonts,amsthm,amssymb,xspace,bm}

\newcommand{\newref}[2][]{\hyperref[#2]{#1~\ref*{#2}}}
\renewcommand{\eqref}[1]{\hyperref[#1]{(\ref*{#1})}}
\numberwithin{equation}{section}
\newcommand{\aref}[1]{\newref[Appendix]{#1}}
\newcommand{\sref}[1]{\newref[Section]{#1}}
\newcommand{\dref}[1]{\newref[Definition]{#1}}
\newcommand{\tref}[1]{\newref[Theorem]{#1}}
\newcommand{\lref}[1]{\newref[Lemma]{#1}}

\newcommand{\cref}[1]{\newref[Corollary]{#1}}

\newcommand{\eref}[1]{\newref[Equation]{#1}}

\theoremstyle{plain}
\newtheorem{theorem}{Theorem}[section]
\newtheorem{lemma}[theorem]{Lemma}
\newtheorem{claim}[theorem]{Claim}

\newtheorem{corollary}[theorem]{Corollary}

\newtheorem{definition}[theorem]{Definition}
\newtheorem{fact}[theorem]{Fact}
\theoremstyle{definition}

\DeclareMathOperator*{\pr}{\mathsf{Pr}} 

\DeclareMathOperator*{\ex}{\mathbb{E}}
\newcommand{\rgta}{\rightarrow}
\newcommand{\lfta}{\leftarrow}
\newcommand{\iprod}[2]{\langle #1,#2\rangle}
\newcommand{\cdf}{\mathsf{CDF}}

\newcommand{\rn}{\mathbb{R}^n}
\newcommand{\rnn}{\mathbb{R}^{n \times n}}

\newcommand{\reals}{\mathbb{R}}

\newcommand{\dpm}{\{1,-1\}}

\newcommand{\sign}{\mathsf{sign}}
\newcommand{\sgn}{\mathsf{sgn}}
\newcommand{\hwt}{H_{w,\theta}}

\newcommand{\hh}{\mathcal{H}}

\newcommand{\gsn}{\mathcal{N}(0,1)^n}

\newcommand{\eps}{\epsilon}

\newcommand{\note}[1]{\marginpar{\tiny *note in TeX*}}
\newcommand{\ignore}[1]{}

\newcommand{\calD}{{\cal D}}

\renewcommand{\phi}{\varphi}

\newcommand{\poly}{\mathrm{poly}}

\newcommand{\eqdef}{\stackrel{\textrm{def}}{=}}

\title{Pseudorandom Generators for Polynomial Threshold Functions\footnote{A preliminary version of this work appeared in STOC 2010.}}

 \author{Raghu Meka$^*$ \qquad David Zuckerman\thanks{Partially supported by NSF Grants CCF-0634811 and CCF-0916160 and THECB ARP Grant 003658-0113-2007.}\\\\
 {Department of Computer Science, University of Texas at Austin }\\
  {\small {\tt \{raghu,diz\}@cs.utexas.edu}}}
\date{}
\begin{document}
\begin{titlepage}
\maketitle
\thispagestyle{empty}
\begin{abstract}
We study the natural question of constructing pseudorandom generators (PRGs)
for low-degree polynomial threshold functions (PTFs). We give a PRG with seed-length $\log
n/\epsilon^{O(d)}$ fooling degree $d$ PTFs with error at most
$\epsilon$. Previously, no nontrivial constructions were known
even for quadratic threshold functions and constant error $\epsilon$. For the class of degree $1$ threshold functions or halfspaces, previously only PRGs with seedlength $O(\log n \log^2(1/\eps)/\eps^2)$ were known. We improve this dependence on the error parameter and construct PRGs with seedlength $O(\log n + \log^2 (1/\eps))$ that $\eps$-fool halfspaces.\ignore{For the class of degree $1$ threshold functions or
halfspaces, we construct PRGs with much better dependence on the error parameter~$\epsilon$ and obtain a PRG with seed-length $O(\log n + \log^2(1/\epsilon))$.\ignore{the following results.
\begin{itemize}
\item A PRG with seed length $O(\log n \log(1/\epsilon))$ for $\epsilon
  \geq 1/\poly(n)$.
\item A PRG with seed length $O(\log n)$ for $\epsilon \geq 1/\poly(\log n)$.
\end{itemize}}
Previously, only PRGs with seed length $O(\log n\log^2(1/\epsilon)/\epsilon^2)$ were
known for halfspaces.} We also obtain PRGs with similar seed lengths for fooling
halfspaces over the $n$-dimensional unit sphere.

The main theme of our constructions and analysis is the use of {\sl invariance principles} to construct pseudorandom generators. We also introduce the notion of monotone read-once branching programs, which is key to improving the dependence on the error rate $\epsilon$ for halfspaces. These techniques may be of independent interest.
\end{abstract}
\end{titlepage}

\section{Introduction}
Polynomial threshold functions are a fundamental class of functions with many important applications in complexity theory \cite{Beigel1993}, learning theory \cite{KlivansS2004}, quantum complexity theory \cite{BealsBCMW2001}, voting theory \cite{AspnesBFR1994} and more. A polynomial threshold function (PTF) of degree $d$ is a function $f:\dpm^n \rgta \dpm$ of the form $f(x) = \sign(P(x)-\theta)$, where $P:\dpm^n \rgta \reals$ is a multi-linear polynomial of degree $d$. Of particular importance are the class of degree $1$ threshold functions, also known as halfspaces, which have been instrumental in the development of many fundamental tools in learning theory such as perceptrons, support vector machines and boosting. 

Here we address the natural problem of explicitly constructing pseudorandom generators (PRGs) for PTFs. Derandomizing natural complexity classes is a fundamental problem in complexity theory, with several applications outside complexity theory. For instance, PRGs for PTFs facilitate estimating the accuracy of PTF classifiers in machine learning with a small number of deterministic samples; PRGs for spherical caps and PRGs for intersections of halfspaces can help derandomize randomized algorithms such as the Goemans-Williamson Max-Cut algorithm.

In this work, we give the first nontrivial pseudorandom generators for low-degree PTFs. 

\begin{definition}
A function $G:\{0,1\}^r \rgta \dpm^n$ is a PRG with error $\epsilon$ for (or $\epsilon$-fools) PTFs of degree $d$, if
\[ |\,\ex_{x \in_u \dpm^n}[f(x)] - \ex_{y \in_u \{0,1\}^r}[f(G(y))]\,| \leq \epsilon ,\]
for all PTFs $f$ of degree at most $d$. (Here $x \in_u S$ denotes a uniformly random element of $S$.)
\end{definition}
We refer to the parameter $r$ as the seed-length of the generator $G$ and say the generator is explicit if it is computable by a (deterministic) polynomial time algorithm. It can be shown by the probabilistic method that there exist PRGs that $\epsilon$-fool degree $d$ PTFs with {\sl seed length} $r = O(d\log n + \log(1/\epsilon))$ (see \aref{app:nonexplicit}). However, despite their long history, until recently very little was known about explicitly constructing such PRGs, even for the special class of halfspaces. 

In this work, we present a PRG that $\epsilon$-fools degree $d$ PTFs with seed length $\log n/\epsilon^{O(d)}$. Previously, PRGs with seed length $o(n)$ were not known even for degree $2$ PTFs and constant $\epsilon$. 

\begin{theorem}\label{mainptfintro}
For $0 < \epsilon < 1$, there exists an explicit PRG fooling PTFs of degree $d$ with error at most $\epsilon$ and seed length $2^{O(d)}\log n/\epsilon^{8d+3}$. 
\end{theorem}

Independent of our work, Diakonikolas et al.~\cite{DiakonikolasKN10} showed that bounded independence fools degree $2$ PTFs and in particular give a PRG with seed-length $(\log n) \cdot \tilde{O}(1/\epsilon^9)$ for degree $2$ PTFs (here $\tilde{O}$ hides poly-logarithmic factors). In another independent work, Ben-Eliezer et al.~\cite{BenLY09} showed that bounded independence fools certain special classes of PTFs.

For the $d = 1$ case of halfspaces, Diakonikolas et al.~\cite{DGJSV09} constructed PRGs with seed length $O(\log n)$ for constant error rates. PRGs with seed length $O(\log^2 n)$ for halfspaces with polynomially bounded weights follow easily from known results. However, nothing nontrivial was known for general halfspaces, for instance, when $\epsilon = 1/\sqrt{n}$. In this work we construct PRGs with exponentially better dependence on the error parameter $\epsilon$.

\begin{theorem}\label{mainhsintro}
For all constants $c$, $\epsilon \geq 1/n^c$, there exists an explicit PRG fooling halfspaces with error at most $\epsilon$ and seed length $O(\log n + \log^2 (1/\epsilon))$. 
\end{theorem}
\ignore{
\begin{theorem}\label{mainintronz}
For all constants $c$, $\epsilon \geq 1/\log^c n$, there exists an explicit PRG fooling halfspaces with error at most $\epsilon$ and seed length $O(\log n)$. 
\end{theorem}}

We also obtain results similar to the above for spherical caps. The problem of constructing PRGs for spherical caps was brought to our attention by Amir Shpilka; Karnin et al.~\cite{KarninRS2009} were the first to obtain a PRG with similar parameters using different methods. They achieve a seed-length of $(1+o(1))\log n + O(\log^2(1/\epsilon))$.
 \begin{theorem}\label{mainspintro}
There exists a constant $c > 0$ such that for all $\epsilon > c \log n/n^{1/4}$, there exists an explicit PRG fooling spherical caps with error at most $\epsilon$ and seed length $O(\log n + \log^2 (1/\epsilon))$. 
 \end{theorem}
\ignore{
 \begin{theorem}\label{mainspintronz}
 For $\epsilon \geq 1/\poly(\log n)$, there exists an explicit PRG fooling spherical caps with error at most $\epsilon$ and seed length $O(\log n)$. 
 \end{theorem}
}
We briefly summarize the previous constructions for halfspaces.

\begin{enumerate}
\item Halfspaces with polynomially bounded integer weights can be computed by polynomial width read-once branching programs (ROBPs). Thus, the PRGs for ROBPs such as those of Nisan \cite{nisanprg} and Impagliazzo et al.~\cite{inw} fool halfspaces with polynomially bounded integer weights with seed length $O(\log^2 n)$. However, a simple counting argument (\cite{MaassT94}, \cite{Hastad94}) shows that almost all halfspaces have exponentially large weights. 
\item Diakonikolas et al.~\cite{DGJSV09} showed that $k$-wise independent spaces fool halfspaces for $k = O(\log^2(1/\epsilon)/\epsilon^2)$. By using the known efficient constructions of $k$-wise independent spaces they obtain PRGs for halfspaces with seed length $O(\log n\log^2(1/\epsilon)/\epsilon^2)$. 
\item Rabani and Shpilka \cite{RabaniS09} gave explicit constructions of polynomial size {\sl hitting sets} for halfspaces.
\end{enumerate}

The overarching theme behind all our constructions is the use of {\sl invariance principles} to get pseudorandom generators. Broadly speaking, invariance principles for a class of functions say that under mild conditions (typically on the first few moments) the distribution of the functions is essentially invariant for all product distributions. Intuitively, invariance principles could be helpful in constructing pseudorandom generators as we can hope to exploit the invariance with respect to product distributions by replacing a product distribution with a ``smaller product distribution'' that still satisfies the conditions for applying the invariance principle. We believe that the above technique could be helpful for other derandomization problems.

Another aspect of our constructions is what we call the ``monotone trick''. The PRGs for small-width read-once branching programs (ROBP) from the works of Nisan \cite{nisanprg}, Impagliazzo et al.~\cite{inw}, and Nisan and Zuckerman \cite{NZ}, have been a fundamental tool in derandomization with several applications \cite{Sivakumar2002}, \cite{RozenmanV2005}, \cite{GopalanR09}. An important ingredient in our PRG for halfspaces is our observation that any PRG for small-width ROBPs fools arbitrary width ``monotone'' ROBPs. Roughly speaking, we say an ROBP is monotone if there exists an ordering on the nodes in each layer of the program so that the corresponding sets of accepting strings {\sl respect the ordering} (see \dref{bpdef}). We believe that this notion of monotone ROBP is quite natural and combined with the ``monotone trick'' could be useful elsewhere.

The above techniques have recently found other applications that we briefly describe in \sref{sec:applications}. We now give a high level view of our constructions and their analyses. 
\subsection{Outline of Constructions}
Our constructions build mainly on the hitting set construction for halfspaces of Rabani and Shpilka. Although the constructions and analyses are similar in spirit for halfspaces and higher degree PTFs, for clarity, we deal with the two classes separately, at the cost of some repetition. The analysis is simpler for halfspaces and provides intuition for the more complicated analysis for higher degree PTFs.
\subsubsection{PRGs for Halfspaces}
Our first step in constructing PRGs for halfspaces is to use our ``monotone trick'' to show that PRGs for polynomial width read-once branching programs (ROBPs) also fool halfspaces. Previously, PRGs for polynomial width ROBPs were only known to fool halfspaces with polynomially bounded weights. Although the natural simulation of halfspaces by ROBP may require polynomially large width, we note that the resulting ROBP is what we call {\sl monotone} (see \dref{bpdef}). We show that PRGs for polynomial width ROBP fool monotone ROBPs of arbitrary width. 

\begin{theorem}\label{monotonebpintro}
A PRG that $\delta$-fools monotone ROBP of width $\log(4T/\epsilon)$ and length $T$ fools monotone ROBP of arbitrary width and length $T$ with error at most $\epsilon + \delta$.
\end{theorem}

See \tref{monotonebp} for a more formal statement. As a corollary we get the following.

\begin{corollary}\label{bpintro}
For all $\epsilon > 0$, a PRG that $\delta$-fools width $\log (4n/\epsilon)$ and length $n$ ROBPs fools halfspaces on $n$ variables with error at most $\epsilon+\delta$. 
\end{corollary}

The above result already improves on the previous constructions for small $\epsilon$, giving a PRG with seed length $O(\log^2 n)$ for $\epsilon = 1/\poly(n)$. However, the randomness used is $O(\log^2 n)$ even for constant $\epsilon$.

We next improve the dependence of the seed length on the error parameter $\epsilon$ to obtain our main results for fooling halfspaces. Following the approach of Diakonikolas et al.~\cite{DGJSV09} we first construct PRGs fooling {\sl regular} halfspaces. A halfspace with coefficients $(w_1,\ldots,w_n)$ is regular if no coefficient is significantly larger than the others. Such halfspaces are easier to analyze because for regular $w$, the distribution of $\iprod{w}{x}$ with $x$ uniformly distributed in $\dpm^n$ is close to a normal distribution by the Central Limit Theorem. Using a quantitative form of the above statement, the Berry-Ess\'een theorem, we show that a simplified version of the hitting set construction of Rabani and Shpilka gives a PRG fooling {\sl regular} halfspaces.

Having fooled regular halfspaces, we use the structural results on halfspaces of Servedio \cite{Servedio06} and Diakonikolas et al.~\cite{DGJSV09} to fool arbitrary halfspaces. The structural results of Servedio and Diakonikolas et al.~roughly show that either a halfspace is regular or is close to a function depending only on a small number of coordinates. Given this, we proceed by a case analysis as in Diakonikolas et al.: if a halfspace is regular, we use the analysis for regular halfspaces; else, we argue that bounded independence suffices.

The above analysis gives a PRG fooling halfspaces with seed length $O(\log n\log^2(1/\epsilon)/\epsilon^2)$, matching the PRG of Diakonikolas et al.~\cite{DGJSV09}. However, not only is our construction simpler to analyze (for the regular case), but we can also apply our ``monotone trick'' to derandomize the construction. Derandomizing using the PRG for ROBPs of Impagliazzo et al.~\cite{inw} gives \tref{mainhsintro}.

For spherical caps, we give a simpler more direct construction based on our generator for regular halfspaces. We use an idea of Ailon and Chazelle \cite{AilonC06} and the invariance of spherical caps with respect to unitary rotations to convert the case of arbitrary spherical caps to {\sl regular spherical caps}. We defer the details to \sref{sec:sphericalhs}.

\subsubsection{PRGs for PTFs}
We next extend our PRG for halfspaces to fool higher degree polynomial threshold functions. The construction we use to fool PTFs is a natural extension of our {\sl underandomized} PRG for halfspaces. The analysis, though similar in outline, is significantly more complicated and at a high level proceeds as follows. 

As was done for halfspaces we first study the case of regular PTFs. The mainstay of our analysis for regular halfspaces is the Berry-Ess\'een theorem for sums of independent random variables. By using the generalized Berry-Ess\'een type theorem, or {\sl invariance principle}, for low-degree multi-linear polynomials, proved by Mossel et al.~\cite{MosselOO2005}, we extend our analysis for regular halfspaces to regular PTFs. We remark that unlike the case for halfspaces, we cannot use the invariance principle of Mossel et al.~directly, but instead adapt their proof technique for our generator. In particular, we crucially use the fact that most of the arguments of Mossel et al.~work even for distributions with bounded independence. 

We then use structural results for PTFs of Diakonikolas et al.~\cite{DSTW09} and Harsha et al.~\cite{HarshaKM09} that generalize the results of Servedio \cite{Servedio06} and Diakonikolas et al.~\cite{DGJSV09} for halfspaces. Roughly speaking, these results show the following: with at least a constant probability, upon randomly restricting a small number of variables, the resulting restricted PTF is either regular or has high bias. However, we cannot yet use the above observation to do a case analysis as was done for halfspaces; instead, we give a more delicate argument with recursive application of the results on random restrictions.

\subsection{Other Applications}\label{sec:applications}
Gopalan et al.~\cite{GopalanOWZ2009} showed that our generator, when suitably modified, fools arbitrary functions of $d$ halfspaces under product distributions where each coordinate has bounded fourth moment. To $\eps$-fool any size-$s$, depth-$d$ decision tree of halfspaces, their generator uses seed length $O((d \log(ds/\eps) + \log n)\cdot \log(ds/\eps))$. For monotone functions of $k$ halfspaces, their seed length becomes $O((k \log(k/\eps) + \log n)\cdot\log(k/\eps))$. They get better bounds for larger $\eps$; for example, to $1/\poly(\log n)$-fool all monotone functions of $(\log n)/\log\log n$ halfspaces, their generator requires a seed of length just $O(\log n)$.

Building on techniques from this work and a new invariance principle for polytopes, Harsha et al.~\cite{HarshaKM09b} obtained pseudorandom generators that $\epsilon$-fool certain classes of intersections of $k$ halfspaces with seed length $(\log n)\cdot\poly(\log k,1/\epsilon)$. As an application of their results, Harsha et al.~obtained the first deterministic quasi-polynomial time approximate-counting algorithms for a large class of integer programs.

In other subsequent work, Gopalan et al.~\cite{GopalanKM10} used ideas motivated by the monotone trick to give the first deterministic polynomial time, relative error approximate-counting algorithms for knapsack and related problems.\\

We first present our result on fooling arbitrary width {\sl monotone} ROBPs with PRGs for small-width ROBPs. 

\newcommand{\zo}{\{0,1\}}
\newcommand{\zdt}{(\zo^D)^T}
\newcommand{\udt}{\mathcal{U}}
\newcommand{\dbp}{D}
\newcommand{\gos}{s}
\section{PRGs for Monotone ROBPs}
We start with some definitions.
\begin{definition}[ROBP]\label{dfn:robp}
An $(S,\dbp,T)$-branching program $M$ is a layered multi-graph with a layer for each $0 \leq i \leq T$ and at most $2^{S}$ vertices (states) in each layer. The first layer has a single vertex $v_0$ and each vertex in the last layer is labeled with $0$ (rejecting) or $1$ (accepting). For $0 \leq i < T$, a vertex $v$ in layer $i$ has exactly $2^\dbp$ outgoing edges each labeled with an element of $\{0,1\}^\dbp$ and ending at a vertex in layer $i+1$. 
\end{definition}

Note that by definition, an $(S,\dbp,T)$-branching program is read-once. We also use the following notation. Let $M$ be an $(S,\dbp,T)$-branching program and $v$ a vertex in layer $i$ of $M$.
\begin{enumerate}
\item For $z = (z^{i},z^{i+1},\ldots,z^T) \in (\{0,1\}^\dbp)^{T+1-i}$ call $(v,z)$ an {\sl accepting} pair if starting from $v$ and traversing the path with edges labeled $z$ in $M$ leads to an accepting state.
\item For $z \in \zdt$,  let $M(z) = 1$ if $(v_1,z)$ is an accepting pair, and $M(z) = 0$ otherwise.
\item $A_M(v) = \{z: (v,z) \text{ is accepting in $M$}\}$ and $P_M(v)$ is the probability that $(v,z)$ is an accepting pair for $z$ chosen uniformly at random.
\item For brevity, let $\udt$ denote the uniform distribution over $\zdt$.
\end{enumerate}

\begin{definition}
A function $G:\{0,1\}^r \rgta (\{0,1\}^D)^T$ is said to $\epsilon$-fool $(S,\dbp,T)$-branching programs if, for all $(S,\dbp,T)$-branching programs $M$,
\[|\, \pr_{z \lfta \udt}\,[M(z) = 1] - \pr_{y \in_u \{0,1\}^r}\,[M(G(y)) = 1]\,| \leq \epsilon.\]
\end{definition}

Nisan \cite{nisanprg} and Impagliazzo et al.~\cite{inw} gave PRGs that $\delta$-fool $(S,\dbp,T)$-branching programs with seed length $r = O((S+\dbp)\log T+\log(T/\delta)\log T)$. For $T = \poly(S,\dbp)$, the PRG of Nisan and Zuckerman \cite{NZ} fools $(S,\dbp,T)$-branching programs with seed length $r = O(S+\dbp)$. We state the bounds of the generator of Impagliazzo et al.~below.
\begin{theorem}[Impagliazzo et al.~\cite{inw}]\label{th:inwprg}
  There exists an explicit generator $G_{INW}:\zo^r \rgta (\{0,1\}^\dbp)^T$ that $\delta$-fools $(S,\dbp,T)$-branching programs with seed-length $r = O(D + (S+\log(T/\delta)\log T))$.
\end{theorem}

Here we show that the above PRGs in fact fool arbitrary width monotone branching programs as defined below.

\begin{definition}[Monotone ROBP]\label{bpdef}
An $(S,\dbp,T)$-branching program $M$ is said to be monotone if for all $0 \leq i < T$, there exists an ordering $\{v_1 \prec v_2 \prec \ldots \prec v_{l_i}\}$ of the vertices in layer $i$ such that for $1 \leq j < k \leq l_i$, $A_M(v_j) \subseteq A_M(v_k)$. 
\end{definition}

\begin{theorem}\label{monotonebp}
Let $0 < \epsilon < 1$ and $G:\{0,1\}^R \rgta (\{0,1\}^\dbp)^T$ be a PRG that $\delta$-fools monotone $(\log(2T/\epsilon),\dbp,T)$-branching programs. Then $G$ fools monotone $(S,\dbp,T)$-branching programs for arbitrary $S$ with error at most $\epsilon+\delta$. 
\end{theorem}

In particular, for $\delta = 1/\poly(T)$ the above theorem gives a PRG fooling monotone $(S,\dbp,T)$-branching programs with error at most $\delta+\epsilon$ and seed length $O(\dbp + \log(T/\epsilon)\log T)$. Note that the seed length does not depend on the space $S$. Given the above result, \cref{bpintro} follows easily.
\begin{proof}[Proof of \cref{bpintro}]
A halfspace with weight vector $w \in \reals^n$ and threshold $\theta \in \reals$ can be naturally computed by an $(S,1,n)$-branching program $M_{w,\theta}$, for $S$ large enough, by letting the states in layer $i$ correspond to the partial sums $\sum_{j=1}^{i} w_j x_j$. It is easy to check that $M_{w,\theta}$ is monotone. The theorem now follows from \tref{monotonebp}.
\end{proof}

\newcommand{\mdn}{M_{down}}
\newcommand{\mup}{M_{up}}

We now prove \tref{monotonebp}. The proof is based on the simple idea of ``sandwiching'' monotone branching programs between small-width branching programs. To this end, let $M$ be a monotone $(S,\dbp,T)$-branching program and call a pair of $(s,\dbp,T)$-branching programs $(\mdn,\mup)$, $\epsilon$-sandwiching for $M$ if the following hold.
\begin{enumerate}
\item For all $z \in (\{0,1\}^\dbp)^T$, $\mdn(z) \leq M(z) \leq \mup(z)$.
\item $\pr_{z \lfta \udt}[\mup(z) = 1] - \pr_{z \lfta \udt}[\mdn(z) = 1] \leq \epsilon$.
\end{enumerate}
We first show that to fool monotone branching programs it suffices to fool small-width sandwiching programs between which the monotone branching program is sandwiched. We then show that every monotone branching program can be sandwiched between two small-width branching programs.

\begin{lemma}\label{lm:bpn1}
If a PRG $G$ $\delta$-fools $(s,\dbp,T)$-branching programs, and there exist $(s,\dbp,T)$-branching programs $(\mdn,\mup)$ that are $\epsilon$-sandwiching for $M$, then $G$ $(\epsilon+\delta)$-fools $M$.
\end{lemma}
\begin{proof}
  Let $\calD$ denote the output distribution of $G$. Then,
\[ \pr_{z \lfta \udt}[\mdn(z) = 1] \leq \pr_{z \lfta \udt}[M(z) = 1],\;\;\;  \pr_{z \lfta \calD}[M(z) = 1] \leq \pr_{z \lfta \calD}[\mup(z) = 1].\]
Further, since $\calD$ $\delta$-fools $\mup$, 
\[ \pr_{z \lfta \calD}[\mup(z) = 1] \leq  \pr_{z \lfta \udt}[\mup(z) = 1] + \delta.\]
Thus, 
\[ \pr_{z \lfta \calD}[M(z) = 1] - \pr_{z \lfta \udt}[ M(z) = 1] \leq  \pr_{z \lfta \udt}[\mup(z) = 1]  - \pr_{z \lfta \udt}[\mdn(z) = 1] + \delta \leq \epsilon + \delta.\]
By a similar argument with the roles of $\mup,\mdn$ interchanged, we get
\[ | \pr_{z \lfta \calD}[M(z) = 1] - \pr_{z \lfta \udt}[ M(z) = 1] | \leq \epsilon + \delta.\]
\end{proof}

\begin{lemma}\label{lm:bpn2}
  For any monotone $(S,\dbp,T)$-branching program $M$, there exist $(\log(2T/\epsilon),\dbp,T)$-branching programs $(\mdn,\mup)$ that are $\epsilon$-sandwiching for $M$.
\end{lemma}
\begin{proof}
We first set up some notation. For $0 \leq i \leq T$, let the vertices in layer $i$ of $M$ be $V^i = \{v^i_1 \prec v^i_2 \prec \ldots \prec v^i_{l_i}\}$. For $J \subseteq V^i$, let $\min(J), \max(J)$ denote the minimum and maximum elements of $J$ under $\prec$. Call $J \subseteq V^i$ an interval if there exist indices $p \leq q$ such that $J = \{v^i_p, v^i_{p+1}, \ldots, v^i_q\}$. 

For each $1 \leq i \leq T$, partition the vertices of layer $i$ into at most $t_i \leq 2T/\epsilon$ {\sl intervals} $J_1^i, J_2^i, \ldots, J_{t_i}^i$ so that for any interval $J_k^i$ and $v,v' \in J_k^i$, 
\begin{equation}
  \label{eq:2}
|P_M(v) - P_M(v')| \leq \frac{\epsilon}{2T}.  
\end{equation}
Let $s = \log(2T/\epsilon)$ and define an $(s,\dbp,T)$-branching program $\mup$ as follows. The vertices in layer $i$ of $\mup$ are $B^i = \{\max(J^i_1), \max(J^i_2),\ldots,\max(J^i_{t_i})\}$ and the edges are placed by {\sl rounding} the edges of $M$ {\sl upwards} as follows. For $v \in B^i$ suppose there is an edge labeled $z$ between $v$ and a vertex $w \in J = J^{i+1}_k$. Then, we place an edge labeled $z$ between $v$ and $\max(J)$. $\mdn$ is defined similarly by using $\min(J)$ instead of $\max(J)$ as above. We claim that $\mup,\mdn$ are $\epsilon$-sandwiching for $M$. We analyze $\mup$ below; the analysis for $\mdn$ is similar.

\begin{claim}\label{monbp3}
For $0 \leq i \leq T$ and $v \in B^i$, $A_M(v) \subseteq A_{\mup}(v)$. In particular, for any $z$, $M(z) \leq \mup(z)$.
\end{claim}
\begin{proof}
Follows from the monotonicity of $M$.
\end{proof}

\begin{claim}\label{clm:bp2}
  For $0 \leq i \leq T$, and $v \in B^i$, $P_{\mup}(v) - P_{M}(v) \leq (T-i) \frac{\epsilon}{2T}$. In particular, for $z$ chosen uniformly at random, $\pr[\mup(z) = 1] - \pr[M(z) = 1] \leq \epsilon/2$.
\end{claim}
\begin{proof}
  The second part of the claim follows from the first. The proof is by downward induction on $i$. For $i = T$, the statement is true trivially. Now, suppose the claim is true for all $j \geq i + 1$. Let $v \in B^i$ and let $z = (z^{i+1},\bar{z})$ be uniformly chosen from $(\{0,1\}^D)^{T-i}$ with $z^{i+1} \in_u \{0,1\}^D$. Let $\Gamma(v,z^{i+1}) \in J(v,z^{i+1}) = J^{i+1}_k$ for one of the intervals of layer $i+1$. Then, the edge labeled $z^{i+1}$ from $v$ goes to $\max(J(v,z^{i+1}))$ in $\mup$. Now, 
  \begin{align*}
    P_{M}(v) &= \sum_{u \in \{0,1\}^D} \,\pr[z^{i+1} = u] \, P_{M}(\Gamma(v,u))&\\
             &\geq \sum_{u \in \{0,1\}^D} \,\pr[z^{i+1} = u] \, \left(P_M(\max(J(v,u)))- \frac{\epsilon}{2T}\right) & \text{(Equation \eqref{eq:2})}\\
& \geq \sum_{u \in \{0,1\}^D} \,\pr[z^{i+1} = u] \, \left(P_{\mup}(\max(J(v,u))) - \frac{(T-i-1)\epsilon}{2T}-\frac{\epsilon}{2T}\right) &\text{(Induction hypothesis)}\\
&= \sum_{u \in \{0,1\}^D} \,\pr[z^{i+1} = u] \, P_{\mup}(\max(J(v,u))) - \frac{(T-i)\epsilon}{2T} &\\
&= P_{\mup}(v) -  \frac{(T-i)\epsilon}{2T} &\text{(Definition of $\mup$)}.
  \end{align*}
The claim now follows from the above equation and induction.
\end{proof}
\lref{lm:bpn2} now follows from Claims \ref{monbp3}, \ref{clm:bp2} and similar arguments for $\mdn$.

\end{proof}


\newcommand{\greg}{G}
\newcommand{\gzs}{r_0}
\section{Main Generator Construction}
We now describe our main construction $\greg$ that serves as a blueprint for all of our constructions. The generator $\greg$ is essentially a simplification of the hitting set construction for halfspaces by Rabani and Shpilka \cite{RabaniS09}. We use the following building blocks. Unless otherwise stated we shall assume without loss of generality that the parameters $n,t$ are powers of $2$.
\begin{enumerate}
\item A family $\hh = \{h:[n] \rgta [t]\}$ of hash functions that is $\alpha$-pairwise independent. That is, for a fixed $k \in [t]$ and $i \neq j \in [n]$,
\begin{equation}\label{pairwise}
  \pr_{h \in_u \hh}[h(i) = k\, \wedge\, h(j) = k] \leq \frac{1+\alpha}{t^2}.
\end{equation}

Efficient constructions of size $|\mathcal{H}| = O(nt)$ are known for any constant $\alpha$, even $\alpha = 0$ (see, e.g., \cite{CarterW77}).
\item A generator $G_0: \{0,1\}^{\gzs} \rgta \dpm^m$ of a $\delta$-almost $k$-wise independent space over $\dpm^m$. A distribution $\calD$ over $\dpm^m$ is $\delta$-almost $k$-wise independent if, for all $\{i_1,\ldots,i_k \} \subseteq [m]$
\[ \sum_{b_1,\ldots,b_k \in \dpm^k}\, \left|\pr_{x \lfta D}[x_{i_1} = b_1,\ldots,x_{i_k} = b_k] - \frac{1}{2^k}\right| \leq \delta.\]
Efficient generators $G_0$ as above with seed length $\gzs = O(k+\log m+\log(1/\delta))$ are known \cite{NaorN93}. Without loss of generality we also assume that for random $x$ output by $G_0$, $\ex[x_i] = 0$ for all $i \leq n$.
\end{enumerate}
Although efficient constructions of hash families $\hh$ and generators $G_0$ as above are known even for $\alpha = 0$, $\delta = 0$ and constant $k$, we work with small but non-zero $\alpha,\delta$, as we will need the more general objects for our analyses. 

The basic idea behind the generator is as follows. We first use the hash functions to distribute the {\sl coordinates} ($[n]$) into buckets. The purpose of this step is to spread out the ``influences'' of the coordinates across buckets. Then, for each bucket we use an independently chosen sample from a $\delta$-almost $k$-wise independent distribution to generate the bits for the coordinate positions mapped to the bucket. The purpose of this step is, roughly, to ``match the first few moments'' of functions restricted to the coordinates in each bucket. The hope then is to subsequently use invariance principles to show closeness in distribution. 

Fix the error parameter $\epsilon > 0$ and let $t$ at most $\poly(\log(1/\epsilon))/\epsilon^2$ to be chosen later. Let $m = n/t$ (assuming without loss of generality that $t$ divides $n$) and let $\mathcal{H}$ be an $\alpha$-pairwise independent hash family. To avoid some technicalities that can be overcome easily, we assume that every hash function $h \in \mathcal{H}$ is evenly distributed, meaning $\forall h, i \in [t]$, $|\{j: h(j) = i, j \in [n]\}| = n/t$. Let $G_0:\{0,1\}^{\gzs} \rgta \dpm^m$ generate a $\delta$-almost $k$-wise independent space for $\delta \geq  \poly(\epsilon,1/n)$ to be chosen later.

Define $\greg: \mathcal{H} \times (\{0,1\}^{\gzs})^t \rgta \{0,1\}^n$ by 
\begin{equation}\label{genreg}
  \greg(h,z^1,\ldots,z^t) = x, \text{ where $x_{|h^{-1}(i)} = G_0(z^i)$ for $i \in [t]$.}
\end{equation}

We will show that for the parameters $t,\alpha,\delta,k$ and $\hh,G_0$ chosen appropriately, the above generator fools halfspaces as well as degree $d$ PTFs. In particular, we fool progressively stronger classes, from halfspaces to degree $d$ PTFs by choosing $\hh$ and $G_0$ progressively stronger. The table below gives a simplified summary of the results we get for different choices of $\hh,G_0$. We define {\sl balanced} hash functions in \dref{dfn:balance}.
\begin{center}
\begin{tabular}{|c|c|c|}\hline
  Hash Family $\hh$ & Generator $G_0$ & Fooling class\\\hline
Pairwise independent & $4$-wise independent & Regular halfspaces, \tref{regularhs} \\\hline
Pairwise independent, {\sl Balanced} & $\Theta(\log t)$-wise independent & Halfspaces, \tref{mainhs} \\ \hline
Pairwise independent & $4d$-wise independent & Regular degree $d$ PTFs, \tref{regularptf}\\\hline
Pairwise independent, {\sl Balanced} & $\Theta(t)$-wise independent & Degree $d$ PTFs, \tref{mainaptf}.\\\hline
\end{tabular}
  
\end{center}

\section{PRGs for Halfspaces}
In this section we show that for appropriately chosen parameters, $\greg$ fools halfspaces. We first show that $\greg$ fools ``regular'' halfspaces to obtain a PRG with seed length $O(\log n/\epsilon^2)$ for regular halfspaces. We then extend the analysis to arbitrary halfspaces to get a PRG with seed length $O(\log n\log^2(1/\epsilon)/\epsilon^2)$ and apply the monotone trick to prove \tref{mainhsintro}.

In the following let $\hwt:\dpm^n \rgta \dpm$ denote a halfspace $\hwt(x) = \sign(\iprod{w}{x}-\theta)$. Unless stated otherwise, we assume throughout that a halfspace $\hwt$ is normalized, meaning $\|w\| = 1$ (here $\|\cdot\|$ is the $l_2$-norm). We measure distance between real-valued distributions $P,Q$ by 
\[ d(P,Q) = \|\cdf(P)-\cdf(Q)\|_\infty = \sup_{t \in \reals}|\pr_{x \lfta P}[x < t] - \pr_{x \lfta Q}[x < t]|,\]
also known as Kolmogorov-Smirnov distance. In particular, we say two real-valued distributions $P,Q$ are $\varepsilon$-close if $d(P,Q) \leq \varepsilon$. We use the fact that Kolmogorov-Smirnov distance is convex.

\begin{lemma}\label{convexity}
For fixed $Q$, the distance function $d(P,Q)$ defined for probability distributions 
over $\reals$ is a convex function.
\end{lemma}

 For $\sigma > 0$, let $\mathcal{N}(0,\sigma)$ denote the normal distribution with mean $0$ and variance $\sigma^2$. We also assume that $\epsilon > 1/n^{.49}$ as otherwise, \tref{mainhsintro} follows from \cref{bpintro}.

\subsection{PRGs for Regular Halfspaces}\label{sec:regularhs}
As was done in Diakonikolas et al.~we first deal with regular halfspaces. 
\begin{definition}
A vector $w \in \rn$ $\epsilon$-regular if $|w_i| \leq \epsilon \|w\|$ for all $i$. A halfspace $\hwt$ is $\epsilon$-regular if $w$ is $\epsilon$-regular.
\end{definition}

Let $t = 1/\epsilon^2$. We claim that for $\hh$ pairwise independent and $G_0$ generating an almost $4$-wise independent distribution, $\greg$ fools regular halfspaces. Note that the randomness used by $\greg$ in this setting is $O(\log n/\epsilon^2)$. 
\begin{theorem}\label{regularhs}
Let $\hh$ be an $\alpha$-almost pairwise independent family for $\alpha = O(1)$ and let $G_0$ generate a $\delta$-almost $4$-wise independent distribution for $\delta = \epsilon^2/4n^5$. Then, $\greg$ defined by \eref{genreg} fools $\epsilon$-regular halfspaces with error at most $O(\epsilon)$ and seed length $O(\log n/\epsilon^2)$. In particular, for $x \in \dpm^n$ generated from $\greg$ and $\epsilon$-regular $w$ with $\|w\| = 1$, the distribution of $\iprod{w}{x}$ is $O(\epsilon)$-close to $\mathcal{N}(0,1)$. 
\end{theorem}

To prove the theorem we will need the Berry-Ess\'een theorem, which gives a quantitative form of the central limit theorem and can be seen as an {\sl invariance principle} for halfspaces.

\begin{theorem}[Theorem 1, XVI.5, \cite{Fellerbook}, \cite{Shevtsova2007}]\label{bet}
Let $Y_1,\ldots,Y_t$ be independent random variables with $E[Y_i] = 0$, $\sum_i E[Y_i^2] = \sigma^2$, $\sum_i E[|Y_i|^3] \leq \rho$. Let $F(.)$ denote the cdf of the random variable $S_n = (Y_1 + \ldots Y_n)/\sigma$, and $\Phi(.)$ denote the cdf of the normal distribution $\mathcal{N}(0,1)$. Then, 
\[ \|F - \Phi\|_\infty = \sup_{z} | F(z) - \Phi(z)| \leq \frac{\rho}{\sigma^3}.\]
\end{theorem}

\begin{corollary}\label{betc}
  Let $Y_1,\ldots,Y_t$ be independent random variables with $E[Y_i] = 0$, $\sum_i E[Y_i^2] = \sigma^2$, $\sum_i E[|Y_i|^4] \leq \rho_4$. Let $F(.)$ denote the cdf of the random variable $S_n = (Y_1 + \ldots Y_n)/\sigma$, and $\Phi(.)$ denote the cdf of the normal distribution $\mathcal{N}(0,1)$. Then, 
\[ \|F - \Phi\|_\infty = \sup_{z} | F(z) - \Phi(z)| \leq \frac{\sqrt{\rho_4}}{\sigma^2}.\]
\end{corollary}
\begin{proof}
  For $1 \leq i \leq n$, by Cauchy-Schwarz, $E[|Y_i|^3] \leq \sqrt{E[Y_i^2]} \cdot \sqrt{E[Y_i^4]}$. Therefore, 
\[ \sum_i E[|Y_i|^3] \leq \sum_i \sqrt{E[Y_i^2]} \cdot \sqrt{E[Y_i^4]} \leq \left(\sum_i E[Y_i^2]\right)^{1/2} \left(\sum_i E[Y_i^4]\right)^{1/2}.\]
The claim now follows from \tref{bet}.
\end{proof}

\begin{lemma}\label{epsgaussian}
For $\epsilon$-regular $w$ with $\|w\|=1$ and $x \in_u \dpm^n$, the distribution of $\iprod{w}{x}$ is $\epsilon$-close to $\mathcal{N}(0,1)$. 
\end{lemma}
\begin{proof}
Let $Y_i = w_i x_i$. Then, $\sum_i \ex[Y_i^2] = 1$ and $\sum_i \ex[Y_i^4] = \sum_i w_i^4 \leq \epsilon^2$. The lemma now follows from \newref[Corollary]{betc}.
\end{proof}

The following lemma says that for a pairwise-independent family of hash functions $\hh$ and $w \in \rn$, the weight of the coefficients is {\sl almost equidistributed} among the buckets.
\begin{lemma}\label{hashing}
Let $\mathcal{H}$ be an $\alpha$-almost pairwise independent family of hash functions from $[n]$ to $[t]$. For $\epsilon$-regular $w$ with $\|w\|=1$, $\sum_{i=1}^t \ex[\|w_{h^{-1}(i)}\|^4] \leq (1+\alpha) \epsilon^2 + \frac{1+\alpha}{t}$.
\end{lemma}
\begin{proof}
Fix $i \in [t]$. For $1 \leq j \leq n$, let $X_j$ be the indicator variable that is $1$ if $h(j) = i$ and $0$ otherwise. Then, $\ex[\|w_{h^{-1}(i)}\|^2] = 1/t$ and 
\[ \|w_{h^{-1}(i)}\|^4 = \left(\sum_{j=1}^n (X_j w_j)^2\right)^2 = \sum_{j=1}^n X_j^4 w_j^4 + \sum_{j\neq k} X_j^2 X_k^2 w_j^2 w_k^2.\]

Now, $\ex[X_j^4] \leq (1+\alpha)/t$ and for $j \neq k$, $\ex[X_j^2 X_k^2] \leq (1+\alpha)/t^2$. Thus, taking expectations of the above equation,
\begin{align*}\label{}
\ex[\|w_{h^{-1}(i)}\|^4] &\leq \frac{1+\alpha}{t}\sum_j w_j^4 + \frac{1+\alpha}{t^2} \sum_{j\neq k} w_j^2 w_k^2\\
&\leq \frac{1+\alpha}{t}(\max_i |w_i|^2) + \frac{1+\alpha}{t^2}\\
&\leq \frac{(1+\alpha)\,\epsilon^2}{t} + \frac{1+\alpha}{t^2}.
\end{align*}
The lemma follows by summing over all $i\in[t]$.
\end{proof}

\begin{proof}[Proof of \tref{regularhs}]

Fix a hash function $h \in \mathcal{H}$. Let $w^i = w_{|h^{-1}(i)}$ for $i \in [t]$. Then, 
\[ \iprod{w}{\greg(h,z)} = \sum_{i=1}^t \iprod{w^i}{G_0(z^i)}.\]

Let random variables $Y_i^h \equiv Y_i \equiv \iprod{w^i}{G_0(z^i)}$ and $Y^h = Y_1 + \ldots + Y_t$. Then, $\ex[Y_i] = 0$ and since $G_0(z^i)$ is $\delta$-almost $4$-wise independent, $|\ex[Y_i^2] - \|w^i\|^2| \leq \delta n^2$. Further, for $1 \leq i \leq t$,
\[ \ex_{x \in_u \dpm^m}[\,\iprod{w^i}{x}^4\,] = \sum_{j=1}^m (w^i_j)^4 + 3 \sum_{p\neq q \in [m]} (w^i_p)^2 (w^i_q)^2 \leq 3 \|w^i\|^4.\]

Since, the above equation depends only on the first four moments of random variable $x$ and $G_0(Z^i)$ is $\delta$-almost $4$-wise independent, it follows that $\ex[Y_i^4] \leq 3 \|w^i\|^4 + \delta n^4$. Thus, $\sum_i \ex[Y_i^2] \geq 1 - \delta n^2t \geq 1/2$ and $\sum_{i=1}^t \ex[Y_i^4] \leq 3\sum_{i=1}^t \|w^i\|^4 + \delta n^5$. Let $\rho_h = \sum_i \|w^i\|^4$. Then, by \newref[Corollary]{betc}, since $\delta \leq \epsilon^2/4n^5$,  for a fixed $h$ the distribution of $Y^h$ is $(\sqrt{3\rho_h}+\epsilon)$-close to $\mathcal{N}(0,1)$. 

Observe that for random $h,z$ the distribution of $Y = \iprod{w}{\greg(h,z)}$ is a convex-combination of the distributions of $Y^h$ for $h \in \mathcal{H}$. Thus, from \lref{convexity}, the distribution of $Y$ is $O(\ex [\sqrt{\rho_h}]+\epsilon)$-close to $\mathcal{N}(0,1)$. Now, by Cauchy-Schwarz $\ex[\sqrt{\rho_h}] \leq \sqrt{\ex[\rho_h]}$. Further, since $w$ is $\epsilon$-regular and $t = 1/\epsilon^2$, it follows from \lref{hashing} that $\ex[\rho_h] = \sum_i \ex[\|w^i\|^4] = \sum_i \ex[\|w_{h^{-1}(i)}\|^4] \leq 2(1+\alpha)\epsilon^2$. Thus, the distribution of $Y$ is $O(\epsilon)$-close to $\mathcal{N}(0,1)$. The theorem now follows from combining this with \lref{epsgaussian}.
\end{proof}


\subsection{PRGs for Arbitrary Halfspaces}\label{sec:arbiths}
We now study arbitrary halfspaces and show that the generator $\greg$ fools arbitrary halfspaces if the family of hash functions $\hh$ and generator $G_0$ satisfy certain stronger properties. We use the following structural result on halfspaces that follows from the results of Servedio \cite{Servedio06} and Diakonikolas et al.~\cite{DGJSV09}.
\begin{theorem}\label{structure}
Let $\hwt$ be a halfspace with $w_1 \geq \ldots \geq w_n$, $\sum w_i^2 = 1$. There exists $K = K(\epsilon) = O(\log^2(1/\epsilon)/\epsilon^2)$ such that one of the following two conditions holds. 
\begin{enumerate}
\item  $w^K = (w_{K(\epsilon)+1},\ldots,w_n)$ is $\epsilon$-regular.
\item Let  $w' = (w_1,\ldots,w_{K(\epsilon)})$ and let $H_{w',\theta}(x) = \sgn(\sum_{i=1}^K w_i x_i - \theta)$. Then, 
\begin{equation}\label{struct1}
  |\pr_{x \leftarrow \calD}[\hwt(x) \neq \mathsf{H}_{w',\theta}(x)] | \leq 2\epsilon,
\end{equation}
where $\calD$ is any distribution satisfying the following conditions for $x \lfta \calD$.
\begin{enumerate}
\item The distribution of $(x_1,\ldots,x_{K})$ is $\epsilon$-close to uniform.
\item With probability at least $1-\epsilon$ over the choice of $(x_1,\ldots,x_K)$, the distribution of $(x_{K+1},\ldots,x_n)$ conditioned on $(x_1,\ldots,x_{K})$ is $(1/n^2)$-almost pairwise independent.
\end{enumerate}
In particular, for distributions $\calD$ as above
\begin{equation}\label{struct2}
  |\,\ex_{x \leftarrow \calD}[\hwt(x)] - \ex_{x \leftarrow \calD}[\mathsf{H}_{w',\theta}(x)]\,|\leq 2\epsilon. 
\end{equation}
\end{enumerate}
\end{theorem}
Servedio and Diakonikolas et al.~show the above result when $\calD$ is the uniform distribution. However, their arguments extend straightforwardly to any distribution $\calD$ as above.   

Given the above theorem, we use a case analysis to analyze $\greg$. If the first condition of the theorem above holds, we use the results of the previous section, \tref{regularhs}, showing that $\greg$ fools regular halfspaces. If the second condition holds, we argue that for $x$ distributed as the output of the generator, the distribution of $(x_1,\ldots,x_{K(\epsilon)})$ is $O(\epsilon)$-close to uniform. 

Let $t = K(\epsilon)$. We need the family of hash functions $\hh:[n] \rgta [t]$ in the construction of $\greg$ to be {\sl balanced} along with being $\alpha$-pairwise independent as in Equation \eqref{pairwise}. Intuitively, a hash family is balanced if with high probability the maximum size of a bucket is small. 
\begin{definition}[Balanced Hash Functions]\label{dfn:balance}
A family of hash functions $\hh = \{h:[n] \rgta [t]$ is $(K,L,\beta)$-balanced if for any $S \subseteq [n]$, $|S| \leq K$, 
\begin{equation}\label{balance}
  \pr_{h \in_u \hh}[\; \max_{j \in [t]}\,(|h^{-1}(j) \cap S|) \geq L\;] \leq \beta .
\end{equation} 
\end{definition}

We use the following construction of balanced hash families due to Lovett et al. \cite{LovettRTV09}. 
\ignore{It is known that a family of $l$-wise independent hash functions is balanced with appropriate parameters. For completeness, we include a proof of the following result in Appendix. The distinction based on $\epsilon$ below will be useful later when derandomizing $\greg$ for halfspaces.}
\begin{theorem}[See Lemma 2.12 in \cite{LovettRTV09}]\label{hashfamily}
Let $t = \log(1/\epsilon)/\epsilon^2$ and $K = K(\epsilon)$ as in \tref{structure}. Then, there exists a $(K,O(\log(1/\epsilon)),1/t^2)$-balanced hash family $\hh:[n] \rgta [t]$ that is also pairwise independent with $|\hh| = \exp(O(\log n + \log^2(1/\epsilon)))$. Moreover, $\hh$ is efficiently samplable.\ignore{
\begin{enumerate}
\item For any constant $c \geq 0$, $\epsilon \geq 1/n^c$, $L = O(\log t) = O(\log (1/\epsilon))$ and $|\hh| = n^{O(\log t)}$. 
\item For any constant $c \geq 0$, $\epsilon \geq 1/(\log n)^c$, $L = O(\log n)$ and $|\hh| = \poly(n)$. 
\end{enumerate}}
\end{theorem}

Let $m = n/t$ and fix $L$ to be one of $O(\log t), O(\log n)$. We also need the generator $G_0:\{0,1\}^{\gzs}\rgta \dpm^m$ to be exactly $4$-wise independent and $\delta$-almost $(L+4)$-wise independent for $\delta = \epsilon^3/tn^5$. Generators $G_0$ as above with $\gzs = O(\log n + \log(1/\delta) + L) = O(\log (n/\epsilon))$ are known \cite{NaorN93}.

 We now show that with $\hh,G_0$ as above, $\greg$ fools halfspaces with error $O(\epsilon)$. The randomness used by the generator is $\log|\hh| + \gzs t = O(\log n \log^2(1/\epsilon)/\epsilon^2)$ and matches the randomness used in the results of Diakonikolas et al.~\cite{DGJSV09}.
\begin{theorem}\label{mainhs}
With $\hh,G_0$ chosen as above, $\greg$ defined by Equation \eqref{genreg} fools halfspaces with error at most $O(\epsilon)$ and seed length $O(\log n \log^2(1/\epsilon)/\epsilon^2)$. 
\end{theorem}
\begin{proof}
Let $\hwt$ be a halfspace and without loss of generality suppose that $w_1 \geq \ldots \geq w_n$ and $\sum_i w_i^2 = 1$. Let $S = \{1,\ldots,K(\epsilon)\}$. Call a hash function {\sl $S$-good} if for all $j \in [t]$, $|S_j| = |S \cap h^{-1}(j)| \leq L$. From \dref{dfn:balance}, a random hash function $h \in_u \hh$ is $S$-good with probability at least $1-1/t^2$. Recall that $\greg(h,z^1,\ldots,z^t) = x$, where $x_{|h^{-1}(j)} = G_0(z^j)$ for $j \in [t]$. Let $\calD$ denote the distribution of the output of $\greg$ and let $x \lfta \calD$.
\begin{claim}\label{clm:arbiths}
  Given an $S$-good hash function $h$, the distribution of $x_{|S}$ is $\epsilon$-close to uniform. Moreover, with probability at least $1-\epsilon$ over the random choices of $x_{|S}$, the distribution of $x$ in the coordinates not in $S$ conditioned on $x_{|S}$ is $(\epsilon^2/4n^5)$-almost $4$-wise independent.
\end{claim}
\begin{proof}
Fix an $S$-good hash function $h$. Since $z^1,\ldots,z^t$ are chosen independently, given the hash function $h$, $x|_{S_1},\ldots,x|_{S_t}$ are independent of each other. Moreover, since the output of $G_0$ is $\delta$-almost $(L+4)$-wise independent and $|S_j| \leq L$ for all $j \in [t]$, $x|_{S_j}$ is $\delta$-close to uniform for all $j \in [t]$. It follows that given an $S$-good hash function $h$, $x|_{S}$ is $(t\delta)$-close to uniform. Further, by a similar argument, for any set $I \subseteq [n]\setminus S$ with $|I| = 4$, the distribution of $x_{|(S\cup I)}$ is $(t\delta)$-close to uniform. It follows that, with probability at least $1-\epsilon$, the distribution of $x_{|I}$ conditioned on $x_{|S}$ is $(t\delta/\epsilon)$-close to uniform. The claim now follows from the above observations and noting that $t\delta = \epsilon^3/4n^5$.
\end{proof}

We can now prove the theorem by a case analysis. Suppose that the weight vector $w$ satisfies condition (2) of \tref{structure}. Observe that from the above claim, $\calD$ satisfies the conditions of \tref{structure} (2). Let $\mathsf{H}_{w_{|S},\theta}(x) = \sgn(\iprod{w_{|S}}{x_{|S}} - \theta)$. Then, from Equation \eqref{struct2}, 
\begin{align*}\label{}
  |\,\ex_{x \leftarrow U_n}[\hwt(x)] - \ex_{x \leftarrow U_n}[\mathsf{H}_{w_{|S},\theta}(x)]\,| \leq 2\epsilon, \\
  |\,\ex_{x \leftarrow \calD}[\hwt(x)] - \ex_{x \leftarrow \calD}[\mathsf{H}_{w_{|S},\theta}(x)]\,| \leq  2\epsilon. 
\end{align*}
Moreover, since the distribution of $x_{|S}$ is $\epsilon$-close to uniform under $\calD$ and $\mathsf{H}_{w_{|S},\theta}(x)$ only depends on $x_{|S}$, 
\[ |\ex_{x \leftarrow U_n}[\mathsf{H}_{w_{|S},\theta}(x)] - \ex_{x \leftarrow \calD}[\mathsf{H}_{w_{|S},\theta}(x)]|\leq \epsilon.\]
Combining the above three equations, we get that
\[ |\ex_{x \leftarrow U_n}[\hwt(x)] - \ex_{x \leftarrow \calD}[\hwt(x)]| \leq 5 \epsilon,\]
and thus $\greg$ fools halfspace $\hwt$ with error at most $5\epsilon$. 
 
Now suppose that condition (1) of \tref{structure} holds and $w_{\bar{S}} = (w_{K(\epsilon)+1},\ldots,w_n)$ is $\epsilon$-regular. Fix an assignment to the variables $x_{|S} = u_{|S}$ and let $x_{\bar{S}} = (x_{k+1},\ldots,x_n)$ and $H_{u}(x_{k+1},\ldots,x_n) = \mathsf{sgn}( \iprod{w_{\bar{S}}}{x_{\bar{S}}} - \theta_u)$, where $\theta_u =  \theta - \iprod{w_{|S}}{x_{|S}}$. We will argue that  with probability at least $1-\epsilon$, conditioned on the values of $x_{|S}$, the output of $\greg$ fools the $\epsilon$-regular halfspace $H_u$ with error $O(\epsilon)$. Given the last statement it follows that $\calD$ fools the halfspace $\hwt$ with error $O(\epsilon)$ since the distribution of $x_{|S}$ under $\calD$ is $\epsilon$-close to uniform. 

Since $\hh$ is a family of pairwise independent hash functions and a random hash function $h \in_u \hh$ is $S$-good with probability at least $1 - 1/t^2$, even when conditioned on being $S$-good, a random hash function $h \in_u \hh$ is $\alpha$-pairwise independent for $\alpha = 1$. Further, from \newref[Claim]{clm:arbiths}, conditioned on the hash function $h$ being $S$-good, with probability at least $1-\epsilon$, even conditioned on $x_{|S}$,  the distribution of $x_{|[n]\setminus S}$ is $(\epsilon^2/4n^5)$-almost $4$-wise independent. Thus, we can apply \tref{regularhs}\footnote{Though \tref{regularhs} was stated for $t = 1/\epsilon^2$, the same argument works for all $t \geq 1/\epsilon^2$.} showing that with probability at least $1-\epsilon$, conditioned on the values of $x_{|S}$, the output of $\greg$ fools $H_u$ with error $O(\epsilon)$.
\end{proof}


\newcommand{\gderand}{G_{\mathsf{D}}}
\subsection{Derandomizing $\greg$}\label{sec:derandhs}
We now derandomize the generator from the previous section and prove \tref{mainhsintro}. The derandomization is motivated by the fact that for a fixed hash function $h$ and $w \in \rn, \theta \in \reals$,  $\sgn(\,\iprod{w}{\greg(h,z^1,\ldots,z^t)}-\theta\,)$ can be computed by a monotone ROBP with $t$ layers. Given this observation, by \tref{monotonebp}, we can use PRGs for small-width ROBP to generate $z^1,\ldots,z^t$ instead of generating them independently as before. 

Let $\gzs,t,m,\hh,G_0$ be set as in the context of \tref{mainhs}. Let $s_0 = \log(2t/\epsilon ) = O(\log(1/\epsilon))$ and let $G_{BP}: \{0,1\}^r \rgta (\{0,1\}^s)^t$ be a PRG fooling $(s_0,\gzs,t)$-branching programs with error $\delta$. Define $\gderand: \hh \times \{0,1\}^r \rgta \dpm^n$ by 
\begin{equation}\label{gnisan}
  \gderand(h,y) = \greg(h,G_{BP}(y)).
\end{equation}
The randomness used by the above generator is $\log|\hh| + r$. We claim that $\gderand$ fools halfspaces with error at most $O(\epsilon + \delta)$. 
\begin{theorem}\label{derandhs}
$\gderand$ fools halfspaces with error $O(\epsilon + \delta)$. 
\end{theorem}
\begin{proof}
Fix a halfspace $\hwt$ and without loss of generality (see \cite{LewisC67} for instance) suppose that $w_1,\ldots,w_n,\theta$ are integers. Let $N = \sum_j |w_j|+|\theta|$. Observe that for any $x \in \dpm^n$, $\iprod{w}{x}-\theta \in \{-N,-N+1,\ldots,0,\ldots,N\}$. Fix a hash function $h \in \hh$. We define a $(\log (2N+1),\gzs,t)$-branching program $M_{h,w}$ that for $z = (z^1,\ldots,z^t) \in (\{0,1\}^{\gzs})^t$ computes $\iprod{w}{\greg(h,z)}$. 

For $i \in [t]$, let $w^i = w_{|h^{-1}(i)}$. Then, for $z = (z^1,\ldots,z^t) \in (\{0,1\}^{\gzs})^t$,  by definition of $\greg$ in \eref{genreg}, 
\[ \iprod{w}{\greg(h,z^1,\ldots,z^t)} = \sum_{i=1}^t \iprod{w^i}{G_0(z^i)}.\]
Define a space-bounded machine $M_{h,w}$ as follows. For each $0 \leq i \leq t$, put $N$ nodes in layer $i$ with labels $1,\ldots,N$. The vertices in layer $i$ correspond to the partial sums $Z_i = \sum_{l=1}^i \iprod{w^l}{G_0(z^l)}$.  Note that all partial sums $Z_i$ lie in $\{-N,-N+1,\ldots,N\}$. Now, given the partial sum $Z_i$ there are $2^{\gzs}$ possible values for $Z_{i+1}$ ranging in $\{Z_i + \iprod{w^{i+1}}{G_0(z)}: z \in \{0,1\}^{\gzs}\}$. We add $2^{\gzs}$ edges correspondingly. Finally, label all vertices in the final layer corresponding to values less than $\theta$ as rejecting and label all other vertices as accepting states. 

It follows from the definition of $M_{h,w}$ that $M_{h,w}$ is monotone and for $z = (z^1,\ldots,z^t) \in (\{0,1\}^{\gzs})^t$, $M_{h,w}(z)$ is an accepting state if and only if $\sgn(\sum_i\iprod{w^i}{G_0(z^i)}-\theta) = \hwt(\greg(h,z))= 1$. Thus, from \tref{monotonebp}, for a fixed $h \in \hh$,
\[| \pr_{z \in_u (\{0,1\}^{\gzs})^t}\,[\hwt(\greg(h,z))= 1] - \pr_{y \in_u \{0,1\}^r}\,[\hwt(\greg(h,G_{BP}(y)))= 1] | \leq \delta + \epsilon.\]

The theorem now follows from the above equation and \tref{mainhs}.
\end{proof}

By choosing the hash family $\hh$ from \tref{hashfamily} and using the PRG of Impagliazzo et al. we get our main result for fooling halfspaces.

\begin{proof}[Proof of \tref{mainhsintro}]
Choose $G_{BP}$ in the above theorem to be the PRG of Impagliazzo et al.~\cite{inw}. To $\epsilon$-fool $(S,D,T)$-ROBPs, the generator of Impagliazzo et al., \tref{th:inwprg}, has a seed-length of $O(D + (S+\log(1/\epsilon))\log T)$. Thus, the seed-length of $G_{BP}$ is $r = O(r_0 + (s_0+\log(1/\epsilon)) \log t) = O(\log n + \log^2(1/\epsilon))$. The theorem follows by choosing the hash family $\hh$ as in \tref{hashfamily}.
\end{proof}
\ignore{
\begin{proof}[Proof of \tref{mainintronz}]
Choose $G_{BP}$ in \tref{derandhs} to be the PRG of Nisan and Zuckerman \cite{NZ}. Since, $\epsilon = 1/\poly\log n$ and $t = 1/\epsilon^2$, the seed-length of $G_{BP}$ is $O(\log n)$. The theorem follows by choosing the hash family $\hh$ to be of size $\poly(n)$ as in \tref{hashfamily} (2).
\end{proof}}


\section{PRGs for Polynomial Threshold Functions}
We now extend our results from the previous sections to construct PRGs for degree $d$ PTFs. We set the parameters of $\greg$ as in \tref{mainhs}, with the main difference being that we take $G_0$ to generate a $k$-wise independent space for $k = O(\log^2(1/\epsilon)/\epsilon^{O(d)} + 4d)$ instead of $O(\log^2(1/\epsilon)/\epsilon^2)$ as was done for fooling halfspaces. The analysis of the construction is, however, more complicated and proceeds as follows.
\begin{enumerate}
\item We first use the invariance principle of Mossel et al.~\cite{MosselOO2005} to deal with {\sl regular} PTFs.
\item We then use the structural results on random restrictions of PTFs of Diakonikolas et al.~\cite{DSTW09} and Harsha et al.~\cite{HarshaKM09} to reduce the case of fooling arbitrary PTFs to that of fooling {\sl regular} PTFs and functions depending only on a few variables.
\end{enumerate}

We carry out the first step above by an extension of the hybrid argument of Mossel et al. where we replace blocks of variables instead of single variables as done by Mossel et al. For this part of the analysis, we also need the {\sl anti-concentration} results of Carbery and Wright \cite{CarberyW2001} for low-degree polynomials over Gaussian distributions.

The second step relies on properties of random restrictions of PTFs similar in spirit to those in \tref{structure} for halfspaces. Roughly speaking, we use the following results. There exists a set $S \subseteq [n]$ of at most $L = 1/\epsilon^{O(d)}$ variables such that for a random restriction of these variables, with probability at least $\Omega(1)$ one of the following happens.
\begin{enumerate}
\item The resulting PTF on the variables in $[n]/S$ is $\epsilon$-regular.
\item The resulting PTF on the variables in $[n]/S$ has high bias.
\end{enumerate}

We then finish the analysis by recursively applying the above claim to show that a generator fooling regular PTFs and having bounded independence also fools arbitrary PTFs.

\subsection{PRGs for Regular PTFs}
Here we extend our result for fooling regular halfspaces, \tref{regularhs}, to {\sl regular} PTFs.
\begin{definition}
Let $P(u_1,\ldots,u_n) = \sum_{I} \alpha_I \prod_{i \in I} u_i$ be a multi-linear polynomial of degree $d$. Let $\|P\|_2^2 = \sum_I \alpha_I^2$ and the influence of $i$'th coordinate $\tau_i(P) = \sum_{I \ni i} \alpha_I^2$. We say $P$ is $\epsilon$-regular if 
\[ \sum_i \tau_i(P)^2 \leq \epsilon^2 \|P\|_2^2.\]
We say a polynomial threshold function $f(x) = \sgn(P(x) - \theta)$ is $\epsilon$-regular if $P$ is $\epsilon$-regular.
\end{definition}
Unless stated otherwise, we will assume throughout that $P$ is normalized with $\|P\|_2^2 = 1$. Fix $d > 0$. Let $t = 1/\epsilon^2, m = n/t$ and let $\hh$ be an $\alpha$-pairwise independent family as in \tref{regularhs}. We assume $G_0:\{0,1\}^{\gzs} \rgta \dpm^m$ generates a $4d$-wise independent space, generalizing the assumption of $4$-wise independence used for fooling regular halfspaces. 
\begin{theorem}\label{regularptf}
Let $\hh$ be an $\alpha$-pairwise independent family for $\alpha = O(1)$ and let $G_0$ generate a $4d$-wise independent distribution. Then, $\greg$ defined by Equation \eqref{genreg} fools $\epsilon$-regular PTFs of degree at most $d$ with error at most $O(d\epsilon^{2/(4d+1)})$.
\end{theorem}

We first prove some useful lemmas. The first lemma is simple.
\begin{lemma}\label{influence}
For a multi-linear polynomial $P$ of degree $d$ with $\|P\| = 1$, $\sum_j \tau_j(P) \leq d$.
\end{lemma}
The following lemma generalizes \lref{hashing} and says that for pairwise independent hash functions and regular polynomials, the total influence is {\sl almost equidistributed} among the buckets.
\begin{lemma}\label{hashingptf}
Let $\mathcal{H} = \{h:[n] \rgta [t]\}$ be a $\alpha$-pairwise independent family of hash functions. Let $P$ be a multi-linear polynomial of degree $d$ with coefficients $(\alpha_J)_{J\subseteq [n]}$ and $\|P\| \leq 1$. For $h \in \mathcal{H}$ let
\[ \tau(h,i) = \sum_{J \cap h^{-1}(i) \neq \emptyset} \alpha_J^2.\]
Then, for $h \in_u \mathcal{H}$
\begin{equation}\label{hashing1}
  \ex_{h}\left[\, \sum_{i=1}^t \tau(h,i)^2\,\right] \leq (1+\alpha)\,\sum_{j=1}^n \tau_j(P)^2 + \frac{(1+\alpha)d^2}{t}.
\end{equation}
\end{lemma}
\begin{proof}
Fix $i \in [t]$ and for $1 \leq j \leq n$, let $X_j$ be the indicator variable that is $1$ if $h(j) = i$ and $0$ otherwise. For brevity, let $\tau_j = \tau_j(P)$ for $j \in [n]$. Now,
\begin{align*}\label{}
  \tau(h,i)= \sum_{J \cap h^{-1}(i) \neq \emptyset} \alpha_J^2 &= \sum_{J} \alpha_J^2\, (\vee_{j \in J} X_j) \\
&\leq \sum_{J} \alpha_J^2 \left( \sum_{j \in J} X_j\right)\\
&= \sum_j X_j \sum_{J: J\ni j} \alpha_J^2 \\
&= \sum_{j} X_j \tau_j.
\end{align*}

Thus,
\[ \tau(h,i)^2 \leq \left(\sum_{j=1}^n X_j \tau_j \right)^2 = \sum_j X_j^2 \tau_j^2 + \sum_{j\neq k} X_j X_k \tau_j \tau_k.\]

Note that $\ex[X_j] \leq (1+\alpha)/t$ and for $j \neq k$, $\ex[X_j X_k] \leq (1+\alpha)/t^2$. Thus, 
\begin{align*}\label{}
\ex[\, \tau(h,i)^2\,] &\leq \frac{1+\alpha}{t}\sum_j \tau_j^2 + \sum_{j\neq k} \tau_j \tau_k \frac{1+\alpha}{t^2}\\
&\leq \frac{1+\alpha}{t}\sum_j \tau_j^2 + \frac{1+\alpha}{t^2} (\sum_j \tau_j)^2 .
\end{align*}

The lemma follows by using \lref{influence} and summing over all $i\in[t]$.
\end{proof}

We also use $(2,4)$-hypercontractivity for degree $d$ polynomials, the anti-concentration bounds for polynomials over log-concave distributions due to Carbery and Wright \cite{CarberyW2001}, and the invariance principle of Mossel et al \cite{MosselOO2005}. We state the relevant results below.

\begin{lemma}[$(2,4)$-hypercontractivity]\label{hypercon}
If $Q,R$ are degree $d$ multilinear polynomials, then for $X \in_u \dpm^n$,
\[ \ex_X\,[Q^2\cdot R^2] \leq 9^d \cdot \ex_X[Q^2]\cdot \ex_X[R^2].\]
In particular, $\ex[Q^4] \leq 9^d \cdot \ex[Q^2]^2$.
\end{lemma}

The following is a special case of Theorem 8 of Carbery-Wright \cite{CarberyW2001} (in their notation, set $q = 2d$ and the distribution $\mu$ to be $\mathcal{N}(0,1)^n$).
\begin{theorem}[Carbery-Wright]\label{cw}
There exists an absolute constant $C$ such that for any multi-linear polynomial $P$ of degree at most $d$ with $\|P\|=1$ and any interval $I \subseteq \reals$ of length $\alpha > 0$, 
\begin{equation*}\label{cweq}
  \pr_{\overline{X}\leftarrow \gsn}[ P(\overline{X}) \in I] \leq C d\,\alpha^{1/d}.
\end{equation*}
\end{theorem}

We use the following structural result of Mossel et al.~\cite{MosselOO2005} that reduces the problem of fooling threshold functions to that of fooling certain {\sl nice} functions which are easier to analyze. 
\begin{definition}
  A function $\psi: \reals \rgta \reals$ is $B$-nice, if $\psi$ is smooth and $|\psi^{''''}(t)| \leq B$ for all $t \in \reals$.
\end{definition}

\begin{lemma}[Mossel et al.]\label{nicetoerr}
Let $X,Y$ be two real-valued random variables such that the following hold.
\begin{enumerate}
\item For any interval $I\subseteq \reals$ of length at most $\alpha$, $\pr[\, X \in I\,] \leq C \alpha^{1/d}$, where $C$ is a constant independent of $\alpha$. 
\item For all $1$-nice functions $\psi$, $|E[\psi(X)] - E[\psi(Y)]| \leq \epsilon^2$.
\end{enumerate}
Then, for all $t > 0$, $|\,\pr[X > t]  - \pr[Y>t]\,| \leq 2 C\, \epsilon^{2/(4d+1)}$.
\end{lemma}

The following theorem is a restatement of the main result of Mossel et al.~who obtain the bound $O(d\, 9^d\, \max_i \tau_i(P)))$ instead of the one below. However, their arguments extend straightforwardly to the following.
\begin{theorem}[Mossel et al.]\label{invariance}
Let $P$ be a multi-linear polynomial of degree at most $d$ with $\|P\| = 1$, $\overline{X} \leftarrow \gsn$ and $\overline{Y} \in_u \dpm^n$. Then, for any $1$-nice function $\psi$, 
\[ |\,\ex[\psi(P(\overline{X}))] - \ex[\psi(P(\overline{Y}))]\,| \leq \frac{9^d}{12}\, \sum_i \tau_i(P)^2 . \]
\end{theorem}

We first prove \tref{regularptf}, assuming the following lemma which says that the generator $G$ fools nice functions of regular polynomials.

\begin{lemma}\label{foolnice}
Let $P$ be an $\epsilon$-regular multi-linear polynomial of degree at most $d$ with $\|P\| = 1$. Let $\overline{Y} \in_u \dpm^n$ and $\overline{Z}$ be distributed as the output of $\greg$. Then, for any $1$-nice function $\psi$, 
\[ |\,E[\psi(P(\overline{Y}))] - E[\psi(P(\overline{Z}))]\,| \leq \frac{1+\alpha}{6}\,d^2\,9^d\,\epsilon^2\]
\end{lemma}

\begin{proof}[Proof of \tref{regularptf}]
Let $P$ be an $\epsilon$-regular polynomial of degree at most $d$ and let $\overline{X} \leftarrow \gsn$. Let $X, Y,Z$ be real-valued random variables defined by $X = P(\overline{X})$, $Y = P(\overline{Y})$ and $Z = P(\overline{Z})$. Then, by \tref{invariance} and \lref{foolnice}, for any $1$-nice function $\psi$, 
\[ |E[\psi(X)] - E[\psi(Y)]| \leq \frac{9^d}{12}\epsilon^2,\;\;\; |E[\psi(Y)] - E[\psi(Z)]| \leq \frac{(1+\alpha)\,d^2\, 9^d\,\epsilon^2}{6}.\]
Hence, 
\[ |E[\psi(X)] - E[\psi(Z)]| = O(d^2 \,9^d\,\epsilon^2).\]
Further, by \tref{cw}, for any interval $I \subseteq \reals$ of length at most $\alpha$, $\pr[\, X \in I\,] = O(\,d\,\alpha^{1/d}\,)$. Therefore, we can apply, \lref{nicetoerr} to $X,Y$ and $X,Z$ to get
\[ |\pr[X > t]  - \pr[Y>t]| = O(d \,\epsilon^{2/(4d+1)}),\;\;\; |\pr[X > t]  - \pr[Z>t]| = O(d\,\epsilon^{2/(4d+1)}).\]
Thus,
\[ |\pr[Y > t]  - \pr[Z >t]| = O(d\,\epsilon^{2/(4d+1)}).\]
\end{proof}

\begin{proof}[Proof of \lref{foolnice}]
Fix a hash function $h \in \hh$. Let $Z_1,\ldots,Z_t$ be $t$ independent samples generated from the $4d$-wise independent space. Let $Y_1,\ldots,Y_t$ be $t$ independent samples chosen uniformly from $\dpm^m$. We will prove the claim via a hybrid argument where we replace the blocks $Y_1,\ldots,Y_t$ with $Z_1,\ldots,Z_t$ progressively.

For $0 \leq i \leq t$, let $X^i$ be the distribution with $X^i_{|h^{-1}(j)} = Z_j$ for $1 \leq j \leq i$ and $X^i_{|h^{-1}(j)} = Y_j$ for $i < j \leq t$. Then, for a fixed hash function $h$, $X^0$ is uniformly distributed over $\dpm^n$ and $X^t$ is distributed as the output of the generator. For $i \in [t]$, let $\tau(h,i)$ be the influence of the $i$'th bucket under $h$, 
\[ \tau(h,i) = \sum_{J \cap h^{-1}(i) \neq \emptyset} \alpha_J^2.\]

\begin{claim}\label{hybridstep}
For $1 \leq i \leq t$, 
\[ |\ex[\psi(P(X^i))] - \ex[\psi(P(X^{i-1}))]| \leq \frac{9^d}{12}\,\tau(h,i)^2.\]
\end{claim}
We will use the following form of the classical Taylor series.
\begin{fact}\label{fct:taylor}
  For any $1$-nice function $\psi:\reals \rgta \reals$, $\alpha,\beta \in \reals$
\[    \psi(\alpha+\beta) = \psi(\alpha) + \psi'(\alpha) \beta + \frac{\psi''(\alpha)}{2} \beta^2 + \frac{\psi'''(\alpha)}{6}\beta^3 + err(\alpha,\beta),\]
where $|err(\alpha,\beta)| \leq \beta^4/24$.
\end{fact}
\begin{proof}
Let $I = h^{-1}(i)$ be the variables that have been changed from $X^{i-1}$ to $X^{i}$. Without loss of generality suppose that $I = \{1,\ldots,m\}$. Let 
\[ P(u_1,\ldots,u_n) = R(u_{m+1},\ldots,u_n) + \sum_{J: J \cap [m] \neq \emptyset} \alpha_J \,\left(\prod_{j \in J} u_j \right),\]
where $R(\;)$ is a multi-linear polynomial of degree at most $d$. Let $S(u_1,\ldots,u_m,u_{m+1},\ldots,u_n)$ denote the degree $d$ multi-linear polynomial given by the second term in the above expression.\\

Observe that $X^{i-1},X^i$ agree on coordinates not in $[m]$. Let $X^{i} = (Z_{1},\ldots,Z_{m},X_{m+1},\ldots,X_n) = (Z,X)$ and $X^{i-1} = (Y_{1},\ldots,Y_{m},X_{m+1},\ldots,X_n) = (Y,X)$. 
Then, 
\[ P(X^{i}) = R(X) + S(Z,X), \;\;\; P(X^{i-1}) = R(X) + S(Y,X) .\]

Now, by using the Taylor series expansion, Fact \ref{fct:taylor}, for $\psi$ at $R(X)$,
\begin{multline*}
  \ex[\psi(P(X^i))] - \ex[\psi(P(X^{i-1}))] = \ex[\psi(R + S(Z,X))] - E[\psi(R + S(Y,X))] \\
=\ex[\,\psi(R) + \psi^{'}(R) S(Z,X) + \frac{\psi^{''}(R)}{2}S(Z,X)^2 + \frac{\psi^{'''}(R)}{6}S(Z,X)^3 \pm \{\leq \frac{1}{24}S(Z,X)^4\}\,] - \\
 \ex[\,\psi(R) + \psi^{'}(R) S(Y,X) + \frac{\psi^{''}(R)}{2}S(Y,X)^2 + \frac{\psi^{'''}(R)}{6}S(Y,X)^3 \pm \{\leq \frac{1}{24}S(Y,X)^4\}\,] 
\end{multline*}

Observe that $X,Y,Z$ are independent of one another and are $4d$-wise independent individually. Since $S(\;)$ has degree at most $d$, it follows that for a fixed assignment of the variables $X_{m+1},\ldots,X_n$ in $X$,
\[ \ex[S(Z,X)] = \ex[S(Y,X)],\;\; \ex[S(Z,X)^2] = \ex[S(Y,X)^2],\]
\[ \ex[S(Z,X)^3] = \ex[S(Y,X)^3],\;\; \ex[S(Z,X)^4] = \ex[S(Y,X)^4] .\]
Combining the above equations we get 
\begin{equation}\label{pf1}
|\ex[\psi(P(X^i))] - \ex[\psi(P(X^{i-1}))]| \leq \frac{1}{12} \ex[\, S(Y,X)^4\,] .  
\end{equation}

Now, using the fact that $S(\;)$ is a multi-linear polynomial of degree at most $d$ and since $(Y,X)$ is $4d$-wise independent, $\ex[\,S(Y,X)^4\,] = \ex[\,S(W)^4\,]$, where $W$ is uniformly distributed over $\dpm^n$. Also note that 
\begin{align*}\label{}
  \ex[S(W)^2] &= \ex\left[\, \left(\,\sum_{J: J \cap [m] \neq \emptyset} \alpha_J \,\left(\prod_{j \in J} W_j \right)\,\right)^2\,\right] \\
&=  \sum_{J: J \cap I \neq \emptyset} \alpha_J^2 \\
&= \tau(h,i).
\end{align*}

Therefore, using the $(2,4)$-hypercontractivity inequality, \lref{hypercon}, $\ex[S(W)^4] \leq 9^d\, \ex[S(W)^2]^2$ and Equation \eqref{pf1},
\begin{align*}\label{}
   |\ex[\psi(P(X^i))] - \ex[\psi(P(X^{i-1}))]| &\leq \frac{1}{12} \ex[\, S(Y,X)^4\,] = \frac{1}{12}\ex[\,S(W)^4\,]\\
&\leq \frac{9^d}{12}\,\ex[S(W)^2]^2 = \frac{9^d}{12}\, \tau(h,i)^2.
\end{align*}
\end{proof}
\end{proof}
\begin{proof}[Proof of \lref{foolnice} Continued]
From Claim~\ref{hybridstep}, for a fixed hash function $h$ we have
\[
|\ex[\psi(P(\overline{Y}))] - \ex[\psi(P(\overline{Z}))]| \leq \sum_{i=1}^t |\ex[\psi(P(X^i))] - \ex[\psi(P(X^{i-1}))]| \leq \frac{9^d}{12} \,\sum_{i=1}^t \tau(h,i)^2.
\]

Therefore, for $h \in_u \hh$, using \lref{hashingptf} and $t = 1/\epsilon^2$,
\[ |\ex[\psi(P(\overline{Y}))] - \ex[\psi(P(\overline{Z}))]| \leq  \frac{9^d}{12} \,\ex_h\left[\sum_{i} \tau(h,i)^2\right] = \frac{9^d}{12} \, (1+\alpha)(1+d^2) \epsilon^2 \leq \frac{(1+\alpha)\, d^2\,9^d\,\epsilon^2 }{6}.\] 
\end{proof}

\subsection{Random Restrictions of PTFs}
We use the following results on random restrictions of Diakonikolas et al.~\cite{DSTW09} and Harsha et al.~\cite{HarshaKM09}. We mainly use the exact statements from the work of Harsha et al., as the notion of regular polynomials from Diakonikolas et al.~is slightly different from ours. Specifically, Diakonikolas et al.~define regularity of a polynomial $P$ by bounding $\max_i (\tau_i(P))$, but in our analysis we use the bound of $\sum_i \tau_i(P)^2$. Diakonikolas et al.~have a statement similar to \lref{ptftodt} below; however, we give a simple argument starting from the main lemmas of Harsha et al.~for completeness. 

Fix a polynomial $P$ of degree at most $d$ and suppose that $\tau_1(P) \geq \tau_2(P) \ldots \geq \tau_n(P)$. Let $K(P,\epsilon) = K$ be the least index $i$ such that, 
\[ \tau_{i+1}(P) \leq \epsilon^2 \sum_{l > i} \tau_l(P) .\]
\begin{lemma}[Lemma 5.1 in Harsha et al.~\cite{HarshaKM09}]\label{case1}
The polynomial $P_{x^K}(Y_{k+1},\ldots,Y_n) = P(x_1,\ldots,x_K,Y_{K+1},\ldots,Y_n)$ in variables $Y_{K+1},\ldots,Y_n$ obtained by choosing $x_1,\ldots,x_K \in_u \dpm$ is $c_d \epsilon$-regular with probability at least $\gamma_d$, for some universal constants $c_d,\gamma_d > 0$.
\end{lemma}

\begin{lemma}[Lemma 5.2 in Harsha et al.~\cite{HarshaKM09}]\label{case2}
There exist universal constants $c,c_d,\delta_d > 0$ such that for $K(P,\epsilon) \geq c\log(1/\epsilon)/\epsilon^2 = L$, the following holds for all $\theta \in \reals$. For a random partial assignment $(x_1,\ldots,x_L) \in_u \dpm^L$ with probability at least $\delta_d$ the following happens. There exists $b \in \dpm$ such that 
\begin{equation}
  \label{eq:dt1}
\pr_{(Y_{L+1},\ldots,Y_n) \leftarrow D}\,[\,\sign(P(x_1,x_2,\ldots,x_L,Y_{L+1},\ldots,Y_n)-\theta) \neq b\,] \leq c_d \epsilon,  
\end{equation}
for any $2d$-wise independent distribution $D$ over $\dpm^{n-L}$. 
\end{lemma}
The above lemma is proven by Harsha et al.~when $D$ is the uniform distribution over $\dpm^{n-L}$. However, their argument extends straightforwardly to $2d$-wise independent distributions $D$.

By repeatedly applying the above lemmas, we show that arbitrary low-degree PTFs can be approximated by small depth decision trees in which the leaf nodes either compute a regular PTF or a function with high bias. We first introduce some notation to this end. 

\begin{definition}
A block decision tree $T$  with block-size $L$ is a decision tree with the following properties. Each internal node of the decision tree reads at most $L$ variables. For each leaf node $\rho \in T$, the output upon reaching the leaf node $\rho$ is a function $f_\rho:\dpm^{V_\rho} \rgta \dpm$, where $V_\rho$ is the set of variables not occurring on the path to the node $\rho$. The depth of $T$ is the length of the longest path from the root of $T$ to a leaf in $T$.
\end{definition}

\begin{definition}\label{defgood}
Given a block decision tree $T$ computing a function $f$, we say that a leaf node $\rho \in T$ is {\sl $(\epsilon,d)$-good} if the function $f_\rho$ satisfies one of the following two properties. 
\begin{enumerate}
\item There exists $b \in \dpm$, such that for any $2d$-wise independent distribution $D$ over $\dpm^{V_\rho}$, 
\[ \pr_{Y \leftarrow D}[f_\rho(Y) \neq b] \leq \epsilon.\]
\item $f_\rho$ is a $\epsilon$-regular degree $d$ PTF.
\end{enumerate}
\end{definition}

We now show a lemma on writing low-degree PTFs as a ``decision tree of regular PTFs''.
\begin{lemma}\label{ptftodt}
There exist universal constants $c'_d,c_d''$ such that the following holds for any degree $d$ polynomial $P$ and PTF $f = \sign(P(\,)-\theta)$. There exists a block decision tree $T$ computing $f$ of block-size $L = c'_d \log(1/\epsilon)/\epsilon^2$ and depth at most $c_d'' \log(1/\epsilon)$, such that with probability at least $1-\epsilon$ a uniformly random walk on the tree leads to an $(\epsilon,d)$-good leaf node.
\end{lemma} 
\begin{proof}
  The proof is by recursively applying Lemmas~\ref{case1} and \ref{case2}. Let $c,c_d,\gamma_d,\delta_d$ be constants from the above lemmas. Let $L$ be defined as in \lref{case2} and let $\alpha = \min(\gamma_d,\delta_d)$. For $S \subseteq [n]$ and a partial assignment $y \in \dpm^S$, let $P_{y}:\dpm^{[n]/S} \rgta \reals$ be the degree at most $d$ polynomial defined by $P_{y}(Y) = P(Z)$, where $Z_i = y_i$ for $i \in S$ and $Z_i = Y_i$ for $i \notin S$. Let $L(y) = \min(K(P_{y},\epsilon), L)$ and let $I(y)$ be the $L(y)$ largest influence coordinates in the polynomial $P_{y}$. We now define a block-decision tree computing $f$ inductively.

Let $y_0 = \emptyset$ and let $I_0 = I(y_0)$. The root of the decision tree reads the variables in $I_0$. For $0 \leq q \leq \log_{1/(1-\alpha)}(1/\epsilon)$ suppose that after $q$ steps we are at a node $\beta$ having read the variables in $S(\beta) \subseteq [n]$ and a corresponding partial assignment $y$. Then, if $P_y$ is $c_d \epsilon$-regular or if $P_y$ satisfies Equation \eqref{eq:dt1} we stop. Else, we make another step and read the values of variables in $I(y)$.

For any leaf node $\rho$, let $y(\rho)$ denote the partial assignment that leads to $\rho$. Then the leaf node $\rho$ outputs the function $f_\rho(Y) = \sign(P_{y(\rho)}(Y) - \theta)$. 

It follows from the construction that $T$ is a block-decision tree computing $f$ with block-size $L$ and depth at most $\log_{1/(1-\alpha)}(1/\epsilon)$. Further, for any internal node $\beta \in T$, by Lemmas~\ref{case1},~\ref{case2} at least $\alpha$ fraction of its children are $(c_d\epsilon,d)$-good. Since any leaf node that is not $(c_d\epsilon,d)$-good is at least $\log_{1/(1-\alpha)}(1/\epsilon)$ far away from the root of $T$, it follows that a uniformly random walk on $T$ leads to a $(c_d\epsilon,d)$-good node with probability at least $1-\epsilon$. The lemma now follows.
\end{proof}

\subsection{PRGs for Arbitrary PTFs}
We now study the case of arbitrary degree $d$ PTFs. As was done for halfspaces, we will show that the generator $\greg$ of Equation \eqref{genreg} fools arbitrary PTFs if the family of hash functions $\hh$ and generator $G_0$ satisfy stronger properties. 

Let $t = c_d c_d' \log^2(1/\epsilon)/\epsilon^2$, $m = n/t$, where $c_d,c_d'$ are the constants from \lref{ptftodt}. We use a family of hash functions $\hh:[n] \rgta [t]$ that are $\alpha$-pairwise independent for $\alpha = O(1)$. We choose the generator $G_0:\{0,1\}^{\gzs} \rgta \dpm^m$ to generate a $(t + 4d)$-wise independent space. Generators $G_0$ with $\gzs = O(t \log n)$ are known. We claim that with the above setting of parameter the generator $\greg$ fools all degree $d$ PTFs.

\begin{theorem}\label{mainaptf}
With $\hh,G_0$ chosen as above, $\greg$ defined by Equation \eqref{genreg} fools degree $d$ PTFs with error at most $O(\epsilon^{2/(4d+1)})$ and seed length $O_d(\log n \log^4(1/\epsilon)/\epsilon^4)$. 
\end{theorem}
\newcommand{\df}{\mathcal{D}_{PTF}}
The bound on the seed length of the generator follows directly from the parameter settings. By carefully tracing the constants involved in our calculations and those in the results of Harsha et al. we need, the exact seed length can be shown to be $a^d \log n \log^4(1/\epsilon)/\epsilon^4$ for a universal constant $a$.

Fix a polynomial $P$ of degree $d$ and a PTF $f(x) = \sign(P(x)-\theta)$ and let $T$ denote the block-decision tree computing $f$ as given by \lref{ptftodt}. Let $\df$ denote the output distribution of the generator $\greg$ with parameters set as above. The intuition behind the proof of the theorem is as follows.
\begin{enumerate}
\item As $\df$ has sufficient bounded independence, the distribution on the leaf nodes of $T$ obtained by taking a walk on $T$ according to inputs chosen from $\df$ is the same as the case when inputs are chosen uniformly. In particular, a random walk on $T$ according to $\df$ leads to a $(\epsilon,d)$-good leaf node with high probability.
\item As $\greg$ fools regular PTFs by \tref{regularptf}, $\df$ will fool the function $f_\rho$ computed at a $(\epsilon,d)$-good leaf node. We also need to address the subtle issue that we really need $\df$ to fool a regular PTF $f_\rho$ even when conditioned on reaching a particular leaf node $\rho$.
\end{enumerate}

We first set up some notation. For a leaf node $\rho \in T$, let $U_\rho = [n]\setminus V_\rho$ be the set of variables seen on the path to $\rho$ and let $a_\rho$ be the corresponding assignment of variables in $U_\rho$ that lead to $\rho$. Further, given an assignment $x$, let $\mathrm{Leaf}(x)$ denote the leaf node reached by taking a walk according to $x$ on $T$.

\begin{lemma}\label{dtwalk}
  For any leaf node $\rho$ of $T$, 
\[ \pr_{x \leftarrow \df}[\mathrm{Leaf}(x) = \rho] = \pr_{x \in_u \dpm^n}[\mathrm{Leaf}(x) = \rho].\]
\end{lemma}
\begin{proof}
  Observe that $\df$ is a $t$-wise independent distribution and that for any $\rho$, $|U_\rho| \leq c_d c_d' \log^2(1/\epsilon)/\epsilon^2 = t$. Thus, 
  \begin{align*}
\pr_{x \leftarrow \df}[\mathrm{Leaf}(x) = \rho] &= \pr_{x \leftarrow \df}[x_{|U_\rho} = a_\rho] = \frac{1}{2^{|U_\rho|}} \\
&= \pr_{x \in_u \dpm^n}[x_{|U_\rho} = a_\rho] = \pr_{x \in_u \dpm^n}[\mathrm{Leaf}(x) = \rho].
  \end{align*}
\end{proof}

\begin{lemma}\label{foolgood}
  Fix an $(\epsilon,d)$-good leaf node $\rho$ of $T$. Then,
\[ |\pr_{x \leftarrow \df}[f_\rho(x_{|V_\rho}) = 1\, |\, x_{|U_\rho} = a_\rho] - \pr_{y \leftarrow \dpm^{V_\rho}}[f_\rho(y) = 1] | = O(\epsilon^{2/(4d+1)}).\]
\end{lemma}
\begin{proof}
  We consider two cases depending on which of the two conditions of \dref{defgood} $f_\rho$ satisfies. 

Case (1) - $f_\rho$ has high bias. Note that $\df$ is a $(t+4d)$-wise independent distribution. Since $|U_\rho| \leq t$, it follows that for $x \leftarrow \df$, even conditioned on $x_{|U_\rho} = a_\rho$, the distribution is $2d$-wise independent. The lemma then follows from the fact that for some $b \in \dpm$, $f_\rho$ evaluates to $b$ with high probability.

Case (2) - $f_\rho$ is an $\epsilon$-regular degree $d$ PTF. We deal with this case by using \tref{regularptf}. Let $x = \greg(h,z^1,\ldots,z^t)$ for $h \in_u \hh$, $z^1,\ldots,z^t \in_u \{0,1\}^{\gzs}$, so $x \leftarrow \df$ as in the definition of $\greg$. Let $h_{\rho}: V_\rho \rgta [t]$ be the restriction of a hash function $h$ to indices in $V_\rho$. For brevity, let $x(\rho) = x_{|V_\rho}$ and let $E_\rho$ be the event $x_{|U_\rho} = a_\rho$. We show that the distribution of $x(\rho)$, conditioned on $E_\rho$, satisfies the conditions of \tref{regularptf}.

Observe that conditioning on $E_\rho$ does not change the distribution of the hash function $h \in_u \hh$ because $|U_\rho| \leq t$ and $\df$ is $t$-wise independent. Thus, even when conditioned on $E_\rho$, the hash functions $h_{\rho}$ are almost pairwise independent. For a hash function $h$, $i \in [t]$, let $B_\rho(h,i) = h^{-1}(i)\setminus V_\rho = h_{\rho}^{-1}(i)$. Now, since $G_0$ generates a $(t+4d)$-wise independent distribution, even conditioned on $E_\rho$, for a fixed hash function $h$, the random variables $x(\rho)_{|B_\rho(h,1)},\, x(\rho)_{|B_\rho(h,2)},\,\ldots,\,x(\rho)_{|B_\rho(h,t)}$ are independent of one another. Moreover, each $x(\rho)_{|B_\rho(h,i)}$ is $4d$-wise independent for $i \in [t]$. 

Thus, even conditioned on $E_\rho$, the distribution of $x(\rho)$ satisfies the conditions of \tref{regularptf} and hence fools the regular degree $d$ PTF $f_\rho$ with error at most $O(\epsilon^{2/(4d+1)})$. The lemma now follows.
\end{proof}

\begin{proof}[Proof of \tref{mainaptf}]
Observe that
\[ \pr_{x \leftarrow \dpm^n}[f(x) = 1] = \sum_{\rho \in Leaves(T)}\, \pr_{x \in_u \dpm^n}[x_{|U_\rho} = a_\rho] \cdot \pr_{y \leftarrow \dpm^{V_\rho}}[f_\rho(y) = 1] .\]
Similarly,
  \[ \pr_{x \leftarrow \df}[f(x) = 1] = \sum_{\rho \in Leaves(T)}\, \pr_{x \leftarrow \df}[x_{|U_\rho} = a_\rho] \cdot \pr_{x \leftarrow \df}[f_\rho(x_{|V_\rho}) = 1\, |\, x_{|U_\rho} = a_\rho] .\]
From the above equations and \lref{dtwalk} it follows that
\begin{multline*}
 |\pr_{x \leftarrow \dpm^n}[f(x) = 1] -  \pr_{x \leftarrow \df}[f(x) = 1] | \leq \\\sum_{\rho \in Leaves(T)} \pr_{x \leftarrow \df}[x_{|U_\rho} = a_\rho] \cdot \left|\, \pr_{x \leftarrow \df}[f_\rho(x_{|V_\rho}) = 1\, |\, x_{|U_\rho} = a_\rho] - \pr_{y \leftarrow \dpm^{V_\rho}}[f_\rho(y) = 1] \,\right|  .
\end{multline*}
Now, by \lref{foolgood} for any $(\epsilon,d)$-good leaf $\rho$ the corresponding term on the right hand side of the above equation is $O(\epsilon^{2/(4d+1)})$. Further, from \lref{ptftodt} we know that a random walk ends at a good leaf with probability at least $1-\epsilon$. It follows that
\[  |\pr_{x \leftarrow \dpm^n}[f(x) = 1] -  \pr_{x \leftarrow \df}[f(x) = 1] | \leq \epsilon\, t= O(\epsilon^{2/(4d+1)}).\]
\end{proof}

Our main theorem on fooling degree $d$ PTFs, \tref{mainptfintro}, follows immediately from the above theorem.


\newcommand{\sph}{\mathcal{S}_{n-1}}
\newcommand{\swt}{S_{w,\theta}} 
\section{PRGs for Spherical Caps}\label{sec:sphericalhs}
We now show how to extend the generator for fooling regular halfspaces and its analysis from \sref{sec:regularhs} to get a PRG for spherical caps and prove \tref{mainspintro}.

Let $\mu$ be a discrete distribution (if not, let's suppose we can discretize $\mu$) over a set $U \subseteq \reals$. Also, suppose that for $X \lfta \mu$, $\ex[X] = 0, \ex[X^2] = 1, \ex[|X|^3] = O(1)$. Given such a distribution $\mu$, a natural approach for extending $\greg$ to $\mu$ is to replace the $k$-wise independent space generator $G_0:\zo^r \rgta \dpm^m$ from Equation \eqref{genreg} with a generator $G_\mu:\zo^r \rgta U^m$ that generates a $k$-wise independent space over $U^m$. It follows from the analysis of \sref{sec:regularhs} that for  $G_\mu$ chosen with appropriate parameters, the above generator fools regular halfspaces over $\mu^n$. It then remains to fool non-regular halfspaces over $\mu^n$. It is reasonable to expect that an analysis similar to that in \sref{sec:arbiths} can be applied to $\mu^n$, provided we have analogues of the results of Servedio and Diakonikolas et al., \tref{structure}, for $\mu^n$. 

The above ideas can be used to get a PRG for spherical caps by noting that a) the uniform distribution over the sphere is close to a product of Gaussians (when the test functions are halfspaces) and b) analogues of \tref{structure} for product of Gaussians follow from known {\sl anti-concentration} properties of the univariate Gaussian distribution. Building on the above argument, Gopalan et al.~\cite{GopalanOWZ2009} recently obtained PRGs fooling halfspaces over ``reasonable'' product distributions. Here we take a different approach and give a simpler, more direct construction for spherical caps based on an idea of Ailon and Chazelle \cite{AilonC06} and the invariance of spherical caps with respect to unitary rotations.
\newcommand{\usp}{\mathcal{U}_{sp}}

Let $\sph = \{x\in \rn: \|x\|_2 = 1\}$ denote the $n$-dimensional sphere. By a {\sl spherical cap} $\swt$  we mean the section of $\sph$ cut by a halfspace, i.e., $\swt \eqdef \{x: x \in \sph,\hwt(x) = 1 \}$. 
\begin{definition}
  A function $G:\zo^r \rgta \sph$ is said to $\epsilon$-fool spherical caps if, for all spherical caps $\swt$, 
\[ |\pr_{x \in_u \sph}[ x \in \swt] - \pr_{y \in_u \zo^r}[G(y) \in \swt] | \leq \epsilon.\]
\end{definition}

Note that the uniform distribution over $\sph$, $\usp$, is not a product distribution. We first show that $\usp$ is close to $\mathcal{N}(0,1/\sqrt{n})^n$ when the test functions are halfspaces. 

\begin{lemma}\label{sctonormal}
 There exists a universal constant $C$ such that for any halfspace $\hwt$, 
\[| \pr_{x \lfta \usp}[\hwt(x) = 1] - \pr_{x \lfta \mathcal{N}(0,1/\sqrt{n})^n}[\hwt(x) = 1] |  \leq \frac{C\,\log n}{n^{1/4}}.\]
In particular, for $x \lfta \usp$, the distribution of $\iprod{w}{x}$ is $O(\sqrt{\log n}/n^{1/4})$-close to $\mathcal{N}(0,1/\sqrt{n})$. 
\end{lemma}
\begin{proof}
Observe that for $x \lfta \mathcal{N}(0,1/\sqrt{n})^n$, $x/\|x\|_2$ is distributed uniformly over $\sph$. Thus, 
\[ \pr_{x \in_u \sph}[\hwt(x) = 1] = \pr_{x \lfta \mathcal{N}(0,1/\sqrt{n})^n}[\hwt\left(\frac{x}{\|x\|_2}\right) = 1].\]
Now, for any $x \in \rn$, 
\[ \left|\iprod{w}{x} - \frac{\iprod{w}{x}}{\|x\|_2}\right| = \frac{|\iprod{w}{x}|}{\|x\|_2} \cdot |\|x\|_2 - 1|.\]
Since for $x \lfta \mathcal{N}(0,1/\sqrt{n})$, $\iprod{w}{x}$ is distributed as $\mathcal{N}(0,1/\sqrt{n})$, for some constant $c_1$, 
\[ \pr_{x \lfta \mathcal{N}(0,1/\sqrt{n})^n}\left[\, |\iprod{w}{x}| \geq \frac{c_1 \sqrt{\log n}}{n^{1/2}}\,\right] \leq \frac{1}{n}.\]
Further, by well-known concentration bounds for the norm of a random Gaussian vector (see \cite{LedouxT}, for instance), it follows that for some constant $c_2 > 0$, 
\[ \pr_{x \lfta \mathcal{N}(0,1/\sqrt{n})^n}\left[\,|\|x\|_2 - 1| \geq \frac{c_2 \sqrt{\log n}}{n^{1/4}}\,\right] \leq \frac{1}{n},\]
Combining the above equations we get
\[ \pr_{x \lfta \mathcal{N}(0,1/\sqrt{n})^n}\left[\,\left|\iprod{w}{x} - \frac{\iprod{w}{x}}{\|x\|_2}\right| \geq \frac{c_1 c_2 \log n}{n^{3/4}}\,\right] \leq \frac{2}{n}.\]

Therefore, for $C = c_1 c_2$, 
\begin{align*}
  \pr_{x \lfta \mathcal{N}(0,1/\sqrt{n})^n}\left[\,\hwt\left(\frac{x}{\|x\|_2}\right)  \neq \hwt(x)\,\right] &\leq  \pr_{x \lfta \mathcal{N}(0,1/\sqrt{n})^n}\left[\, |\iprod{w}{x} - \theta| \leq \left|\iprod{w}{x} - \frac{\iprod{w}{x}}{\|x\|_2}\right|\,\right]\\
&\leq \pr_{x \lfta \mathcal{N}(0,1/\sqrt{n})^n}\left[\,|\iprod{w}{x} - \theta| \leq \frac{c_1 c_2 \log n}{n^{3/4}}\,\right] + \frac{2}{n}\\
&\leq \frac{C \log n}{n^{1/4}},
\end{align*}
where the last inequality follows from the fact that $\iprod{w}{x}$ is distributed as $\mathcal{N}(0,1/\sqrt{n})$ and for any interval $I \subseteq \reals$, $\pr_{x \lfta \mathcal{N}(0,1)}[ x \in I] = O(|I|)$.
\end{proof}

Now, by \tref{regularhs}, for $\epsilon$-regular $w$ and $x$ generated from $\greg$ with parameters as in \tref{regularhs}, the distribution of $\iprod{w}{x/\sqrt{n}}$ is $O(\epsilon)$-close to $\mathcal{N}(0,1/\sqrt{n})$. It then follows from the above lemma that  $\greg$  $\epsilon$-fools spherical caps $\swt$ when $w$ is $\epsilon$-regular and $\epsilon \geq C \log n/n^{1/4}$. We now reduce the case of arbitrary spherical caps to {\sl regular spherical caps}.

\newcommand{\calR}{\mathcal{R}}
 Observe that the {\sl volume} of a spherical cap $\swt$ is invariant under rotations: for any unitary matrix $A \in \rnn$ with $A^TA = I_n$, 
\[ \pr_{x \lfta \usp}[x \in \swt] = \pr_{x \lfta \usp}[Ax \in \swt].\]

 We exploit this fact by using a family of rotations $\calR$ of Ailon and Chazelle \cite{AilonC06} which satisfies the property that for any $w \in \rn$ and a random rotation $V \in_u \calR$, $Vw$ is regular with high probability. Let $H \in \rnn$ be the normalized Hadamard matrix such that $H^T H = I_n$ and each entry $H_{ij} \in \{\pm 1/\sqrt{n}\}$. For a vector $x \in \rn$, let $D(x)$ denote the diagonal matrix with diagonal entries given by $x$. Observe that for $x \in \dpm^n$, $HD(x)$ is a unitary matrix. Ailon and Chazelle (essentially) show that for any $w \in \rn$ and $x \in_u \dpm^n$, $HD(x)w$ is $O(\sqrt{\log n}/\sqrt{n})$-regular. We derandomize their construction by showing that similar guarantees hold for $x$ chosen from a $8$-wise independent distribution. 

 \begin{lemma}\label{lm:ac}
For all $w \in \rn$, $\|w\| = 1$, and $x \in \dpm^n$ chosen from an $8$-wise independent distribution the following holds. For $v = HD(x)w$, $\gamma > 0$, 
\[ \pr [ \,\sum_i v_i^4 \geq \frac{\gamma }{n}\,] = O\left(\frac{1}{\gamma^2}\right).\]
 \end{lemma}
 \begin{proof}
   Let random variable $Z = \sum_i v_i^4$. Observe that each $v_i$ is a linear function of $x$ and 
\[ \ex[v_i^2] = \ex[\,(\,\sum_j H_{ij} x_j w_j\,)^2\,] = \sum_j H_{ij}^2 w_j^2 = \frac{1}{n}.\]
Note that since $x$ is $8$-wise independent, we can apply $(2,4)$-hypercontractivity, \lref{hypercon}, to $v_i$. Thus, 
  \[      \ex[Z] = \sum_i \ex[v_i^4] \leq 9 \sum_i \ex[v_i^2]^2 \leq \frac{9}{n}.\]
Similarly, by $(2,4)$-hypercontractivity applied to the quadratics $v_i^2, v_j^2$, 
\[ \ex[Z^2] = \sum_{i,j} \ex[v_i^4 v_j^4] \leq \sum_{i,j} 9^2 \ex[v_i^4] \ex[v_j^4] \leq 9^2 \ex[Z]^2 \leq \frac{9^4}{n^2}.\]
The lemma now follows from the above equation and Markov's inequality applied to $Z^2$.
 \end{proof}

Combining the above lemmas we get the following analogue of \tref{regularhs} for spherical caps. Let $\greg$ be as in \tref{regularhs} and let $\calD$ be a $8$-wise independent distribution over $\dpm^n$. Define $G_{sph}:\dpm^n \times \zo^r \rgta \sph$ by 
\[ G_{sph}(x,y) = \frac{D(x)H^T\greg(y)}{\sqrt{n}}.\]
\begin{theorem}
  For any spherical cap $\swt$ with $\|w\| = 1$ and $\epsilon > C \log n/n^{1/4}$,
\[| \pr_{z \lfta \usp}[\,\iprod{w}{z} \geq \theta\, ] - \pr_{ x \lfta \calD, y \in_u \zo^r}[ \,\iprod{w}{G_{sph}(x,y)} \geq \theta\,]| = O(\epsilon).\]
\end{theorem}
\begin{proof}
By \lref{sctonormal}, for $z \lfta \usp$, $\iprod{w}{z}$ is $O(\epsilon)$-close to $\mathcal{N}(0,1/\sqrt{n})$. Further, by applying \lref{lm:ac} for $\gamma = 1/\sqrt{\epsilon}$, we get that  $v = HD(x)w$ is $\delta$-regular with probability at least $1-O(\epsilon)$ for $\delta = 1/(\sqrt{n}\epsilon^{1/4}) < \epsilon$. Now, by \tref{regularhs} for $v$ $\epsilon$-regular and $y \in_u \zo^r$, the distribution of $\iprod{v}{\greg(y)}$ is $O(\epsilon)$-close to $\mathcal{N}(0,1)$. The theorem now follows from combining the above claims and noting that $\iprod{v}{\greg(y)/\sqrt{n}} = \iprod{w}{G_{sph}(x,y)}$.
\end{proof}

\tref{mainspintro} now follows from the above theorem and derandomizing $\greg$ as done in \sref{sec:derandhs} for proving \tref{mainhsintro}.


\section*{Acknowledgements}
We thank Omer Reingold for allowing us to use his observation improving the seed-length of \tref{mainhsintro} from the conference version. 
The preliminary version of this work appearing in STOC 2010 had a worse seed-length of $O(\log n\log(1/\epsilon))$. However, a minor change in the argument where we use the PRG for small space machines of Impagliazzo et al.~\cite{inw} instead of the PRG of Nisan \cite{nisanprg} in the {\sl monotone trick} leads to the new improved parameters. We thank Amir Shpilka for drawing to our attention the problem of fooling spherical caps and pointing us to the work of Ailon and Chazelle.  We thank Parikshit Gopalan, Prahladh Harsha, Adam Klivans and Ryan O'Donnell for useful discussions and comments. 
{\small
\bibliographystyle{prahladhurl}
\bibliography{references}
}
\appendix
\section{Non-Explicit Bounds}\label{app:nonexplicit}
It is known (\cite{LewisC67}, \cite{RoychowdhurySOK91}) that the number of distinct halfspaces on $n$ bits is at most $2^{n^2}$. One way of extending this bound to degree $d$ PTFs is as follows. It is known that the Fourier coefficients of the first $d+1$ levels of a degree $d$ PTF, also known as the Chow parameters, determine the PTF completely (see \cite{ODonnellS08}). Thus, a PTF $f$ is completely determined by $\mathsf{ChowParam}(f) = (\,\ex[f \cdot \chi_I]: I \subseteq [n], |I| \leq d\,)$, where $\chi_I(x) = \prod_{i \in I} x_i$ denotes the parity over the coordinates in $I$. Observe that for any $I \subseteq [n]$, $\ex[\,f \cdot \chi_I\,] \in \{i/2^n: i \in \mathbb{Z}, |i| \leq 2^n\}$. Therefore, the number of distinct degree $d$ PTFs is at most the number of distinct sequences $\mathsf{ChowParam}(\,)$, which in turn is at most $(2^n)^{n^{d}}$.

The non-explicit bound now follows by observing that any class of boolean functions $\mathcal{F}$ can be fooled with error at most $\epsilon$ by a set of size at most $O(\log(|\mathcal{F}|)/\epsilon^2)$. Thus, degree $d$ PTFs can be fooled by a sample space of size at most $O(n^{d+1}/\epsilon^2)$. 

\end{document}